\documentclass[aip, cha, sd, reprint, groupedaddress, showpacs, floatfix]{revtex4-1}
\usepackage{hyperref}
\usepackage[pdftex]{graphicx}
\usepackage{amsmath,amssymb}
\hypersetup{colorlinks,citecolor=black,filecolor=black,linkcolor=black,urlcolor=black}

\usepackage{empheq}
\usepackage[parfill]{parskip}
\usepackage[active]{srcltx}

\renewcommand{\Re}[1]{\text{Re}{\left(#1\right)}}
\renewcommand{\Im}[1]{\text{Im}{\left(#1\right)}}

\newcommand{\ceil}[1]{{\left\lceil #1 \right\rceil}}
\renewcommand{\d}{\text{d}}
\newcommand{\half}{\frac{1}{2}}
\newcommand{\Subpop}{S}
\renewcommand{\SS}{$\mathcal{SS}$}
\renewcommand{\sp}{\sigma}
\newcommand{\spp}{\tau}
\newcommand{\zangle}{\varphi}
\newcommand{\LL}{{\textsf{LL}}}
\newcommand{\LD}{{\textsf{LD}}}
\newcommand{\DL}{{\textsf{DL}}}
\newcommand{\INC}{{\textsf{INC}}}
\newcommand{\LLLL}{{\textsf{LLLL}}}

\begin{document}

\title{A Two-Frequency-Two-Coupling model of coupled oscillators}

\author{Hyunsuk Hong}
\email{hhong@jbnu.ac.kr}
\affiliation{Department of Physics and Research Institute of Physics and Chemistry, Jeonbuk National University, Jeonju 54896, Korea}

\author{Erik A. Martens}
\email{erik.martens@math.lth.se}
\affiliation{Centre for Mathematical Sciences, Lund University, Box 118, 221 00 Lund, Sweden}
\affiliation{Chair for Network Dynamics, Center for Advancing Electronics Dresden (cfaed) and Institute for Theoretical Physics, TU Dresden, 01062 Dresden, Germany}

\date{\today}

\begin{abstract}
We considered the phase coherence dynamics in a {\it{Two-Frequency and Two-Coupling}} (TFTC) model of coupled oscillators, where coupling strength and natural oscillator frequencies for individual oscillators may assume one of two values (positive/negative). The bimodal distributions for the coupling strengths and frequencies are either {\it{correlated}} or {\it{uncorrelated}}. To study how correlation affects phase coherence, we analyzed the TFTC model by means of numerical simulation and exact dimensional reduction methods allowing to study the collective dynamics in terms of local order parameters~\cite{Watanabe1994,Ott2008}. The competition resulting from distributed coupling strengths and natural frequencies produces nontrivial dynamic states.
For correlated disorder in frequencies and coupling strengths, we found that the entire oscillator population splits into two subpopulations, both phase-locked ({\it{Lock-Lock}}), or one phase-locked and the other drifting ({\it{Lock-Drift}}), where the mean-fields of the subpopulations maintain a constant non-zero phase difference.
For uncorrelated disorder, we found that the oscillator population may split into four phase-locked subpopulations, forming phase-locked pairs, which are either mutually frequency-locked ({\it{Stable Lock-Lock-Lock-Lock}}) or drifting ({\it{Breathing Lock-Lock-Lock-Lock}}), thus resulting in a periodic motion of the global synchronization level.
Finally, we found for both types of disorder that a state of {\it Incoherence} exists; however, for correlated coupling strengths and frequencies, Incoherence is always unstable, whereas it is only neutrally stable for the uncorrelated case.
Numerical simulations performed on the model show good agreement with the analytic predictions.
The simplicity of the model promises that real-world systems can be found which display the dynamics induced by correlated/uncorrelated disorder.
\end{abstract}

\maketitle

\begin{quotation}
The synchronization of oscillators is a ubiquitous phenomenon that manifests itself in a vast range of settings in nature and technology, such as the beating of the heart~\cite{Michaels1987}, circadian clocks in the brain~\cite{Liu1997}, metabolic oscillations in yeast cells~\cite{Dano1999}, and life cycles of phytoplankton~\cite{Massie2010}, pedestrians on a bridge locking their gait~\cite{Strogatz2005}, metronomes on a swing~\cite{Pantaleone2002}, arrays of Josephson junctions~\cite{Wiesenfeld1998}, chemical oscillators~\cite{Kiss2002,Tinsley2012}, electric power grid networks~\cite{Rohden2012} and others~\cite{Strogatz2003}.
Studies have addressed collective dynamics emerging in coupled oscillator networks with properties giving rise to the formation subpopulation structures induced by either heterogeneous frequencies~\cite{Crawford1994,Pazo2009,Martens2009,Pietras2018} or interactions such as coupling strengths~\cite{Abrams2008,Montbrio2004,Wildie2012,Deschle2019} and/or phase-lags~\cite{Martens2016,Bick2018,Choe2016}, with bimodal character. Recently, Hong {\it et al.}~\cite{Hong2016} considered the emergence of collective states including traveling waves in a system with heterogeneous natural frequencies and positive / negative coupling strengths that were correlated with the given natural frequency. 
Here, we simplify this model by considering the case where  both natural frequencies and coupling strengths may assume either positive or negative value, which may be correlated or uncorrelated. Thereby, we compared side-by-side the impact of correlated and uncorrelated disorder and study the collective dynamics based on numerical simulations and dimensional reduction methods~\cite{Bick2020}. The resulting collective dynamics reflects the subpopulation structure imprinted by the natural frequencies/coupling strengths in a nontrivial way, depending on the amount of asymmetry of the disorder. On {the} one hand, the correlated model exhibits dynamic states where one oscillators {belonging to one subpopulation are locked, while oscillators in the other subpopulation are adrift}, or both subpopulations are frequency-locked with constant phase difference; both states display traveling wave motion. On the other hand, for uncorrelated disorder we found a state of incoherent oscillations, and a state where all subpopulations are phase-locked, either with a drifting or constant phase difference; traveling wave motion is however absent. {Our findings corroborate that traveling wave motion results from asymmetry in natural frequencies and coupling strengths with correlated disorder, rather than from disorder without correlation or disorder with non-zero variance.}
\end{quotation}

\section{Introduction}

Synchronization phenomena occur in a large variety of systems in nature and technology, and a wide range of studies used both mathematical and physical models to uncover and understand the dynamics of synchronization~\cite{Kuramoto1984, Strogatz2000, PikovskyBook2001,  Acebron2005}.  To explore the mechanisms behind collective synchronization, the paradigmatic Kuramoto model is a useful tool~\cite{Kuramoto1984}, and its variants capture many features of biological and physical systems in the real world, including pedestrians walking on a bridge~\cite{Strogatz2005}, Josephson junctions~\cite{Wiesenfeld1998}, neural systems~\cite{Sompolinsky1990,Breakspear2010}, metronomes~\cite{Pantaleone2002,MartensThutupalli2013}, lasers~\cite{} and opto-mechanical systems~\cite{Heinrich2011}. For example, the collective synchronization in Kuramoto model has attracted physicists' attention because its governing equations can be related to the {\it XY} model for the spin magnetics, i.e., the Kuramoto model corresponds to an overdamped version of the Hamiltonian dynamics of the {\it XY} model in physics~\cite{Witthaut2014}. The model has also attracted great theoretical attention because of its analytical tractability via the exact low dimensional description of the microscopic dynamics in terms of collective mean-field variables, for a review see~\cite{Bick2020}.

The natural frequencies of the  oscillators in the original Kuramoto model~\cite{Kuramoto1984} are {randomly drawn from a unimodal distribution} function such as the Gaussian one{, while the coupling strength between all oscillators is the same value,} and consequently, frequencies and coupling strengths are uncorrelated. The natural frequencies of the oscillators play two roles. The first is that the frequencies constitute driving forces in the system. The second is that they play the role of {\it{``disorder''}} due to their randomly distributed nature.  This disorder in the oscillator frequencies tends to break synchrony and forces the oscillator phases to run away from each other; conversely, {(positive)} coupling strength is an antagonist to this disorder and enables the oscillators to entrain their phases. 
In Ref.~\cite{Hong2016}, {one of the authors} considered a system where the natural frequencies and the coupling strengths are drawn from random distributions with finite variance. In particular, the authors considered the case where the distributions of the two parameters are symmetrically/asymmetrically correlated with each other and found that the correlation may induce interesting states including traveling waves. In the present study, we considered a minimal model where both natural frequencies and coupling strengths are centered around two distinct values and investigated how correlated/uncorrelated disorder in natural frequencies and coupling strengths affected the collective dynamics of the system. Coupling strengths hereby take on both positive and negative values. 

The motivation for this article is to study the effects of correlated/uncorrelated disorder by simplifying {a previous model}~\cite{Hong2016} such that these effects are analytically tractable. To achieve this, we introduce a \emph{``Two-Frequencies-Two-Coupling (TFTC) model''} where frequencies and coupling strengths may assume either of two values. In particular, we would like to address the following questions:
Previous studies reported~\cite{Hong2016} intriguing dynamic states such as traveling waves induced by correlated disorder; 
can we observe traveling waves despite the simplifications in this model, and what other dynamic states may appear?  

This paper is structured as follows. Sec.~II defines the TFTC model of coupled oscillators, Sec.~III gives a dimensional reduction to this system in terms of macroscopic collective dynamics of based on the theories introduced by Ott/Antonsen~\cite{Ott2008,Ott2009} and Watanabe/Strogatz~\cite{Watanabe1993,Watanabe1994}. In Sec.~IV we carry out a stability and bifurcation analysis for the dynamics resulting from the ``correlated model'', where the natural oscillator frequency and coupling strength are correlated with each other, using numerical simulation and the dimensionally reduced equations. In Sec.~V we carry out a similar analysis for the ``uncorrelated model'', where natural oscillator frequencies and coupling strengths are randomly chosen, using numerical simulation, a self-consistency argument and the dimensionally reduced equations. Finally, Sec.~VI provides a summary and discussion of our results.

\section{Model}
We consider a minimal model of coupled oscillators in which oscillators may assume either of two values for their natural frequencies and their coupling strengths --- hence we refer to  it as the  ``{\it{Two-Frequency and Two-Coupling (TFTC)} }'' model.
The dynamics of the $j=1,\ldots,N$, $N\in \mathbb{N}$, oscillators is described in terms of their phases, $\phi_j \in \mathbb{R}/2\pi\mathbb{Z} \simeq [0, 2\pi)$, representing points on the unit circle $S^1$, and evolve according to  a variant of the Kuramoto model,
\begin{equation}\label{eq:model}
    \frac{d\phi_j}{dt} = \omega_j + \frac{1}{N} \sum_{k=1}^{N} \xi_k \sin(\phi_k-\phi_j), ~~~j= 1,\ldots, N,
\end{equation}
 where the natural frequencies $\omega_j$ are drawn from a bimodal distribution function,
\begin{equation}\label{eq:gw}
    g(\omega) = p \delta(\omega+q\gamma)+q\delta(\omega-p\gamma), 
\end{equation} 
where $\delta$ denotes the Kronecker-delta distribution.    
Thus, oscillators have either a negative frequency, $\omega=-q\gamma$, 
with probability $p$, or a positive frequency, $\omega=p\gamma$, with probability $q:=1-p$.
The parameter $\gamma=|p\gamma-(-q\gamma)|>0$ defines the spacing between the two peaks and, since $\langle\omega\rangle=\int \omega g(\omega) \d \omega=p(-q\gamma)+qp\gamma=0$, the distribution has always zero mean. 
The coupling strength, $\xi_k$, defines the interaction strength between oscillator, $k$, and all other oscillators, $j=1,\ldots,N$, and is  assumed to be either positive or negative. The coupling strengths are drawn from the bimodal distribution function,
\begin{equation}\label{eq:Gamma_xi}
    \Gamma(\xi) = p\delta(\xi-1)+q\delta(\xi+1).
\end{equation}
We may either rescale the coupling strength, $\xi_k$, or frequencies (time), $\omega_k$. Here, we chose to keep the distance between peaks of the coupling strength fixed, while the distance between peaks of the frequencies remains tunable via $\gamma$.
To simplify the problem, we assume that the parameter $p$ is identical in the two distributions given by \eqref{eq:gw} and \eqref{eq:Gamma_xi}. Thus, oscillators either have positive coupling strength ($\xi=1$) with probability $p$, or negative coupling strength ($\xi=-1$) with probability $q=1-p$. 

By choosing the coupling strength and natural frequency according to \eqref{eq:gw} and \eqref{eq:Gamma_xi}, we introduce a certain type of {\it{disorder}} in the system. We consider two model variants:

\paragraph{Correlated model.} We consider the case where the two types of disorders, namely, in natural frequencies and in coupling strengths, are {\it{correlated}} with one another. One may envision various ways to introduce correlation between the two disorders; however, we consider a very simple way of correlating the two distributions of coupling strengths, $ \xi_j $, and frequencies, $\omega_j $. Specifically, we observe that coupling strengths with either $ \xi = +1 $ or $ \xi = -1 $ split the population into two subpopulations,  
$\Subpop_1$ and $\Subpop_2$, containing a number of elements corresponding to integer values near $pN$  and $qN$, respectively. This can be achieved by defining  the subpopulations as follows:
$\Subpop_1:=\{1,\ldots,\iota(p)\}$ and $\Subpop_2:=\{\iota(p)+1,\ldots,{N}\}$  with $\iota(p):=\ceil{p(N-1)}$ for $0<p<1$; $\Subpop_1:=\{\}$ and $\Subpop_2:=\{1,\ldots, N\}$  for $p=0$; and $\Subpop_1:=\{1,\ldots,N\}$ and $\Subpop_2:=\{\}$ for $p=1$. 
Correlation between frequencies and coupling strengths is then invoked by the following rule:
\begin{equation}
    \omega_j  =
    \left\{
    \begin{array}{ll}
    -q\gamma \xi_j , \quad &{\rm for}~j \in \Subpop_1,\\
    -p\gamma \xi_j , \quad &{\rm for}~j \in \Subpop_2.\
    \end{array}
    \right.
    \label{eq:w_xi} 
\end{equation}
This choice for the correlated disorder divides oscillators into two sub-populations, $ \sp = 1,2$, with properties:
\begin{align}\label{eq:correlated}
\begin{split}
    (\xi^{(1)},\omega^{(1)})&=(+1,-q\gamma),\\
    (\xi^{(2)},\omega^{(2)})&=(-1,+p\gamma).\
\end{split}
\end{align}

\paragraph{Uncorrelated model.} The natural frequencies $\omega$, drawn from the distribution $g(\omega)$, and the coupling strengths $
\xi$, drawn from the distribution $\Gamma(\xi)$, are independent from one another. Thus, the uncorrelated model divides oscillators into four sub-populations, $ \sp = 1,2,3,4$, reflecting the two properties assigned to the oscillators:
\begin{align}\label{eq:uncorrelated}
\begin{split}
    (\xi^{(1)},\omega^{(1)})&=(+1,-q\gamma),\\
    (\xi^{(2)},\omega^{(2)})&=(-1,+p\gamma),\\
    (\xi^{(3)},\omega^{(3)})&=(+1,+p\gamma),\\
    (\xi^{(4)},\omega^{(4)})&=(-1,-q\gamma).\
\end{split}
\end{align}

The grouping of properties resulting from these two models imprints a subpopulation structure that allows us to rewrite Eq.~\eqref{eq:model} as follows:
\begin{align}
    \label{eq:model_2subpopulations}
    { \dot \phi_j^{(\sp)}} = \omega^{(\sp)} + \frac{1}{N}\sum_{\spp=1}^{|M|} \xi^{(\tau)}\sum_{ k \in \Subpop_{\spp} }  \sin( \phi_k^{(\tau)} - \phi_j^{(\sp)} ),
\end{align}
where $\phi_j^{(\sp)}$ is the phase of oscillator $j=1,\ldots,|\Subpop_\sigma|$ belonging to subpopulation $\Subpop_\sp$, and $M=2$ or $M=4$ for the correlated and uncorrelated model, respectively. 

\paragraph{Characterization of collective dynamics.}
The collective dynamic behavior observed for the governing equations~\eqref{eq:model} and both models may be characterized by the complex order parameter, 
\begin{equation}\label{eq:Z}
    Z = R e^{i\Psi} = \frac{1}{N}\sum_{k=1}^N e^{i\phi_k}, 
\end{equation}
or the weighted complex order parameter, 
\begin{equation}\label{eq:W}
    W =  S e^{i\Delta } = \frac{1}{N}\sum_{k=1}^N \xi_k e^{i\phi_k}.
\end{equation}
Both order parameters measure the synchronization level in the oscillator population: for incoherent oscillations, phases spread uniformly on the circle such that $R=S=0$; synchronized phase-locked motion can be characterized by $R=1$.

Note that the value of the weighted order parameter, $ S = | W | $, can be smaller than 1 even if phase-locked motion occurs with $ R = 1 $.  Accordingly, a perfectly synchronized/coherent state is characterized by $ R = 1 $ and $ 0 < S < 1 $. By contrast, a state of {\it{``partial synchronization''}} implies that some of the oscillators exhibit synchronized phase-locked motion while others are adrift, and thus the state is characterized by $ R < 1 $; or  an all-frequency locked state with distributed phases.  The state with $ R = 0 $, but $ S > 0 $ is not possible in the current system, as we show further below using the numerical simulations, see also \eqref{eq:W} and \eqref{eq:Z}, and the comments following thereafter.

\section{Dimensional reduction}\label{sec:OttAntonsen}
We restrict our analysis to the case of large systems in the continuum limit, $N\rightarrow\infty$. 
The continuum limit prompts a statistical description in terms of a density function describing the phases of oscillators, $\rho=\rho(\phi,\omega,\xi,t)$, which evolves according to the continuity equation
\begin{align}\label{eq:continuity}
    \frac{\partial}{\partial t}f + \frac{\partial}{\partial \phi}(\rho v)&=0,\
\end{align}
where the velocity is given by
\begin{align}\label{eq:goveq_continuous}
    v &=\omega + \Im{W(t)e^{-i\phi}},\
\end{align}
where the weighted order parameter,
\begin{align}\label{eq:W_continuous}
    W(t) &= \int_{-\infty}^\infty\int_{-\infty}^\infty\int_{-\pi}^\pi\xi' \Gamma(\xi')g(\omega')\rho(\xi',\omega',\phi',t)e^{i\phi'}\d \phi'\d\xi'\d\omega',\
\end{align}
acts as a mean-field forcing on each oscillator.

The Ott-Antonsen method~\cite{Ott2008} formulates a solution for the phase density via Fourier series ansatz,
\begin{align}
    f &= \frac{1}{2\pi}g(\omega)\Gamma(\xi) (1 + f^+ + \bar{f}^+)
\end{align}
with 
\begin{align}
    f^+&=\sum_{k=1}^\infty a(\xi,\omega,t)^k e^{ik \phi}\
\end{align}
where we assume that $f^+$ has an analytic continuation into the lower complex plane. We may recognize this ansatz for the phase density as the Poisson kernel, parameterized by $a=re^{i\phi}$. Geometrically, the Ott-Antonsen manifold defines a two dimensional submanifold in the {infinite-dimensional} space of density functions. Substitution of this ansatz into \eqref{eq:continuity} results in a infinite set of identical ordinary differential equations, the amplitude equations for each mode $e^{ik\phi}$:
\begin{align}\label{eq:amplitude_equations}
    \dot{a}&=-i\omega a + \half (\bar{W}-Wa^2).\
\end{align}
When these amplitude equations are satisfied, the phase density $\rho$ is restricted to the invariant Poisson manifold~\cite{Ott2008}. 

The integro-o.d.e. system defined by \eqref{eq:amplitude_equations} and \eqref{eq:W_continuous} can be further simplified. Substituting the Ott-Antonsen ansatz and carrying out the integral over the phases, we have
\begin{align}\label{eq:W_continuous2}
    W(t) &= \int_{-\infty}^\infty\int_{-\infty}^\infty \xi\, \bar{a}(\xi,\omega,t)\Gamma(\xi)g(\omega)\,\d\omega\,\d\xi,\
\end{align}
and cleverly choosing the distribution functions $g$ and $\Gamma$ allow to take the contour integral in $W$ over the lower complex plane and express $W$ in terms of expressions in $a$~\cite{Ott2008}. For the current models, evaluating the integral in $W$ is particularly simple due to nature of choices for $g$ and $\Gamma$. Here, the particular choices for $g(\omega)$ and $\Gamma(\xi)$ give rise to {the} subpopulation structure explained for the correlated model in \eqref{eq:correlated} and for the uncorrelated model in \eqref{eq:uncorrelated} which also organizes the macroscopic dynamics in terms of local order parameters, $z_\sigma$, as we show next.
The correlated model implies that 
$ g(\omega)\Gamma(\xi) = p\,\delta(\omega+q\gamma)\,\delta(\xi-1)
+ q\,\delta(\omega-q\gamma)\,\delta(\xi+1) $ and we therefore have 
\begin{align}\label{eq:W_correlated}
    W(t) &= p z_1 - q z_2,\
\end{align}
where we defined
\begin{align}\label{eq:z_correlated}
    z_1(t) &:= \bar{a}(\xi=+1, \omega=-q\gamma, t), \nonumber\\
    z_2(t) &:= \bar{a}(\xi=-1, \omega=+p\gamma, t).
\end{align}

Using $g(\omega)$ as defined in Eq.~\eqref{eq:gw} for the uncorrelated model, we obtain
\begin{align}\label{eq:W_uncorrelated}\nonumber
    W(t)&= p\int_{-\infty}^\infty\xi \bar{a}(\xi,-q\gamma,t)\Gamma(\xi)\,\d\xi  \\
    &+q\int_{-\infty}^\infty \xi \bar{a}(\xi,p\gamma,t) \Gamma(\xi) \,\d\xi \nonumber\\
    &=p^2 z_1 - pqz_4 + pq z_3 - q^2z_2.\
\end{align}
where we defined
\begin{align}\label{eq:z_uncorrelated}
    z_1(t) &:= \bar{a}(\xi=+1, \omega=-q\gamma, t), \nonumber\\
    z_2(t) &:= \bar{a}(\xi=-1, \omega=+p\gamma, t), \nonumber\\
    z_3(t) &:= \bar{a}(\xi=+1, \omega=+p\gamma, t), \nonumber\\
    z_4(t) &:= \bar{a}(\xi=-1, \omega=-q\gamma, t). 
\end{align}
Thus, the dynamics of $z_\sp$ are given in closed form by
\begin{align}\label{eq:z_dynamics}
    \dot{z}_\sp &= i \omega^{(\sp)} z_\sigma +\half \left(W - \bar{W} z_\sp^2\right),
\end{align}
for each subpopulation $\sp=1,2$ (correlated model) or $\sp=1,2,3,4$ (uncorrelated model).

Later, we shall use the complex order parameter which in the $N\rightarrow\infty$ limit is defined as
\begin{align}\label{eq:Z_continuous}
    Z(t) &= \int_{-\infty}^\infty\int_{-\infty}^\infty\int_{-\pi}^\pi \Gamma(\xi')g(\omega')\rho(\xi',\omega',\phi',t)e^{i\phi'}\d \phi'\d\xi'\d\omega'.\
\end{align}
For the correlated model, we find
\begin{align}\label{eq:ZZ_correlated}
    Z(t) &=p z_1+qz_2,\
\end{align}
and for the uncorrelated model, 
\begin{align}\label{eq:Z_uncorrelated}
    Z(t) &=p^2 z_1 + pqz_2 + pq z_3+ q^2 z_2.\
\end{align}

We note that \eqref{eq:W_uncorrelated} and \eqref{eq:z_dynamics} may also be obtained from considering the dynamics of oscillator (sub-)populations with identical natural frequencies. Watanabe and Strogatz~\cite{Watanabe1993} showed that the phase space of an oscillator population is foliated by 3-dimensional leafs determined by $N_\sigma-3$ constants of motion $\theta_j^\sp \in\mathbb{R}/2\pi\mathbb{Z}$, $j=1,\ldots,N_\sigma-3$. Dynamics for each subpopulation  $\sigma$  in  \eqref{eq:model_2subpopulations} are constrained to submanifolds of dimension at most three, governed by~\cite{Watanabe1994,Pikovsky2008}, 

\begin{subequations}
 \begin{eqnarray}
    \label{eq:WS_1}
    \dot\rho_\sp &=& \frac{1-\rho^2_\sp}{2} \Re{W e^{-i\Phi_\sp}},\\
    \label{eq:WS_2}
    \dot\Phi_\sp &=& \omega_\sp + \frac{1+\rho^2_\sp}{2\rho_\sp} \Im{W e^{-i\Phi_\sp}}, \\
    \label{eq:WS_3}
    \dot\Theta_\sp &=& \frac{1-\rho^2_\sp}{2\rho_\sp} \Im{W e^{-i\Phi_\sp}}, 
\end{eqnarray}
\end{subequations}
where $\Re{\cdot}$ and $\Im{\cdot}$ represent the real and imaginary parts of a complex number, respectively. Suppose now that the level of synchronization inside each subpopulation $\sp$ is  characterized by the magnitude $0\leq r_\sp\leq 1$ of the local complex order parameter given by
\begin{equation}\label{eq:local_complex_OP}
z_\sp := r_\sp e^{i\zangle_\sp}=\frac{1}{N_\sp}\sum_{k\in \Subpop_\sp} e^{i\phi_k}.
\end{equation}
Assuming that the constants of motion are uniformly distributed, $\theta_j^\sp=2\pi j/N_\sigma$, $j=1,\ldots,N_\sigma-3$,  and that the number of oscillators tends to infinity, $N_\sigma\rightarrow\infty$, one can show that the equalities $r_\sp=\rho_\sp$ and $\varphi_\sp=\Phi_\sigma$ hold~\cite{Pikovsky2011} and the dynamics of $\rho_\sigma=r_\sigma$ and $\Phi_\sigma=\varphi_\sigma$ decouple from the dynamics of $\Theta_\sigma$. 
Using these relations in Eq.~\eqref{eq:WS_1}, recalling that $z_\sp=r_\sp e^{i\varphi_\sp}$ and using that $N_1/N\rightarrow p$ and $N_2/N\rightarrow q$, we obtain from Eqs.~\eqref{eq:WS_1}-\eqref{eq:WS_2} equations identical to \eqref{eq:z_dynamics} describing the dynamics on the Poisson (or Ott-Antonsen) manifold~\cite{Marvel2009}. For a review on dimensional reduction methods developed by Ott/Antonsen and Watanabe/Strogatz and their details, see~\cite{Bick2020} and references therein.

\section{Analysis for correlated disorder}

\subsection{Numerical Simulations\label{sec:correlated_num_sim}}
\begin{figure}[!htp]
    \centering
    \includegraphics[width=0.95\linewidth]{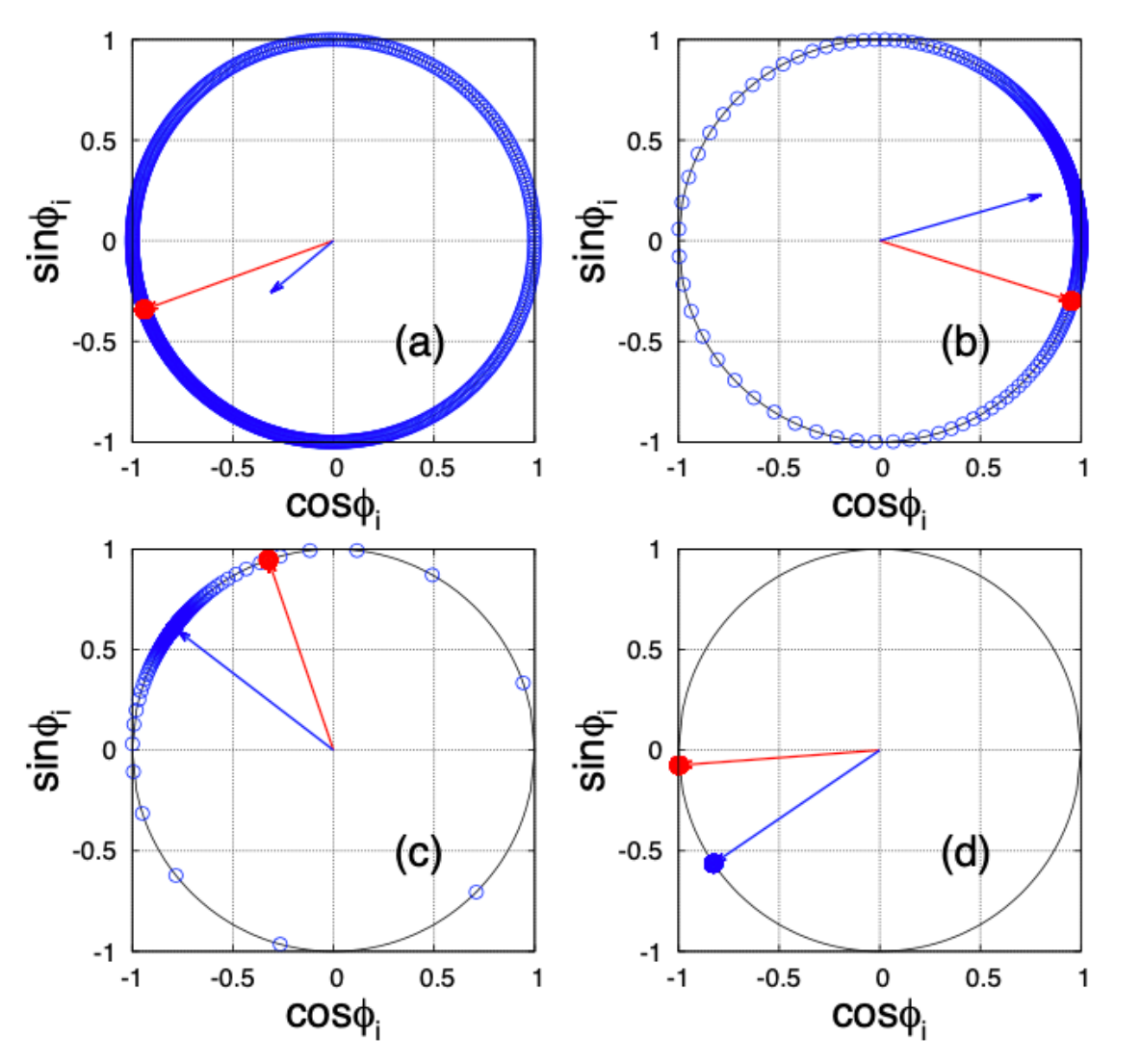}
    \caption{\label{fig:phi_correlated} (Color Online) Snapshot of the phases $\phi_i$ are shown in the plane of $(\cos\phi_i, \sin\phi_i)$ for $N=1000$ oscillators in the correlated model.
    Lock-Drift (\LD) states are observed for $p=0.3$ (a), $p=0.4$ (b), $p=0.54$ (c); the Lock-Lock (\LL) state is observed for $p=0.55 (>p_c)$ in panel (d), where $p_c$ is the critical point where the \LL\ state loses stability and connects to the \LD\ state (see Sec.~\ref{sec:corr_equilibria}).  Red/blue circles represent phases of the oscillators belonging to $\Subpop_1$ / $\Subpop_2$, and red/blue arrows show order parameters $z_1$ and $z_2$, respectively.
}
\end{figure}

We obtained first insights into the possible dynamic behavior for the model with correlated disorder via numerical simulations of Eqs.~\eqref{eq:model}, using a fourth-order Runge-Kutta (RK4) integration scheme with a time step of $\Delta t = 0.01$, over a simulation time of $M_t=2\times 10^5$. {For any given value of $p$, initial phases $\{\phi_i(0)\}$ were randomly drawn from a uniform distribution on the interval $[0, 2\pi)$.} 
Snapshots of asymptotic states of phases at time $t=M_t$ are shown in Fig.~\ref{fig:phi_correlated} for several values of $p$. For all reported values, the oscillator population splits into two subpopulations, where the first, $\Subpop_1$ (red), is formed by oscillators with $\xi^{(1)}=+1$ and $\omega^{(1)}=-q\gamma$, and the second, $\Subpop_2$ (blue), is formed by oscillators with  $\xi^{(2)}=-1$ and $\omega^{(2)}=+p\gamma$.

We found that the system may exhibit at least two states: 
\begin{itemize}
    \item[i)] The {\it Lock-Drift state} (\LD) where oscillators split into two subpopulations, one phase-locked with $r_1=1$ ($\Subpop_1$) and the other ($\Subpop_2$) drifting with $r_2<1$, as shown in panels (a)-(c). The subpopulations are frequency-locked so that their phase difference $\delta:=\phi_2-\phi_1$ remains constant, as shown in the analysis further below.
 
    \item[ii)] The {\it Lock-Lock} (\LL) state where all oscillators split into two phase-locked subpopulations with $r_1=r_2=1$ rotating at a constant frequency with fixed phase distance, $\delta>0$, as shown in panel (d).
\end{itemize}

Note that the {\it Drift-Lock} state, that is, the symmetric {counterpart} of the \LD\ state where the role between the two subpopulations is reversed, is not observed. We can understand this as follows.
Oscillators in subpopulation $\Subpop_1$ with positive coupling strength $\xi^{(1)}=+1$ tend to minimize their phase differences, thus leading to ``phase-locking behavior''. Vice versa, oscillators in subpopulation $\Subpop_2$ with negative coupling strength $\xi^{(2)}=-1$ tend to maximize their phase difference, thus leading to ``drifting'' behavior. Therefore, we observe the \LD\ state where subpopulation $\Subpop_1$ and $\Subpop_2$ assume locked and drifting dynamic behavior, respectively; conversely, the \DL\ state, where the roles of the two subpopulations is reversed, does not emerge, as discussed in Sec.~\ref{sec:corr_absence_L-D_state}.

\begin{figure}[!htp]
    \includegraphics[width=0.95\linewidth]{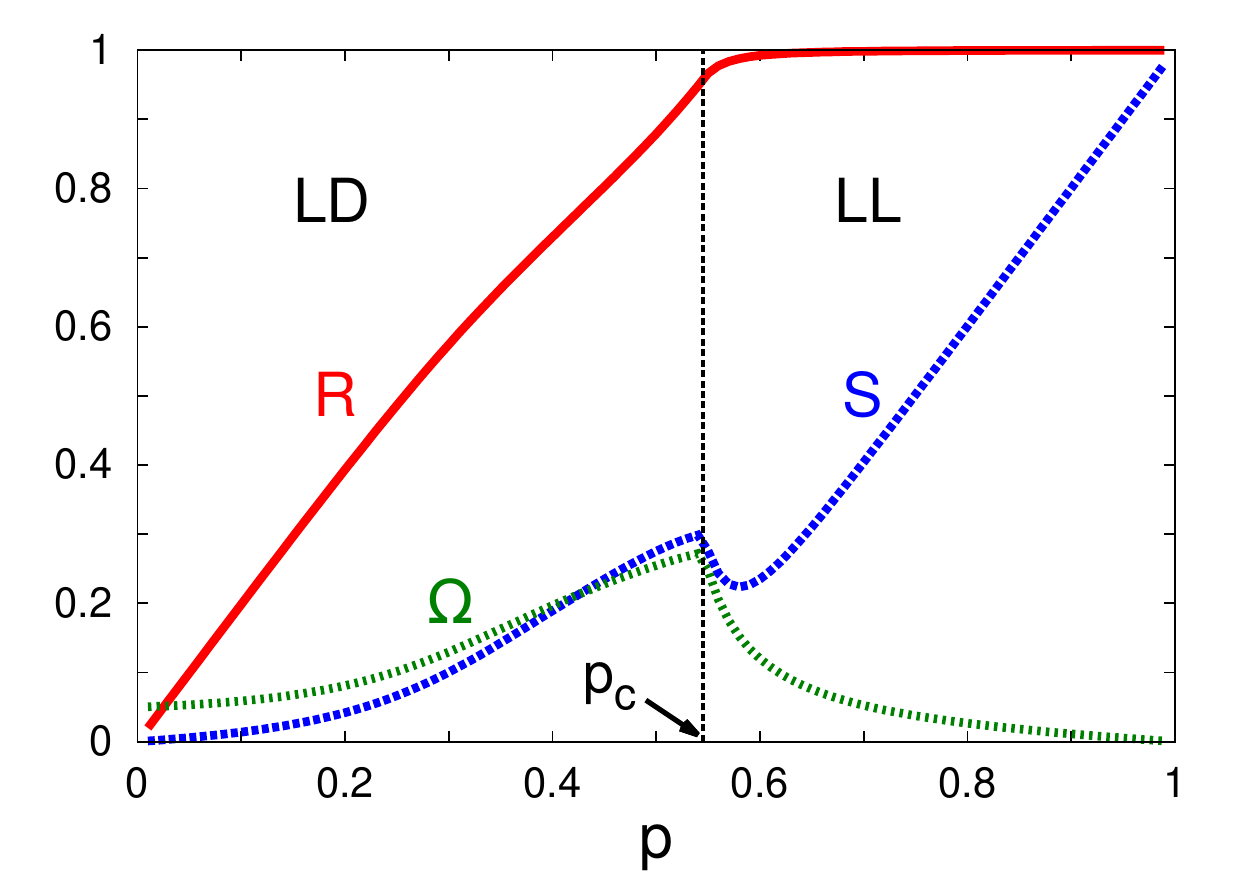}
    \caption{\label{fig:R_S}(Color Online)   Order parameters $R$ (red solid line) and $S$ (blue dashed line) are shown as a function of $p$ while $\gamma=0.05$ is kept fixed,  where $R$ and $S$ are given by Eq.~(\ref{eq:R}) and (\ref{eq:S}), with $r_1$, $r_2$, and $\delta$ shown in Eqs.~(\ref{eq:FixedPoints_LL_delta}) and (\ref{eq:r1r2_smallp}). The label \LD\ represents the Lock-Drift state shown for $p<p_c$, and the label \LL\ is the Lock-Lock state for $p>p_c$, respectively.
    The critical value $p_c$ is given by Eq.~(\ref{eq:pcrit}).
    The dark green dotted line displays the behavior of the wave speed $\Omega$ {(with arbitrary scale)} from Eq.~(\ref{eq:Omega}). 
}
\end{figure}

Furthermore, to gain insight into the ranges of existence of these state, we obtained the phase diagram shown in Fig.~\ref{fig:R_S}, where we measured the time asymptotic behavior of several macroscopic variables while varying $p$ for $\gamma=0.05$ fixed, averaged over the time interval $[M_t/2,M_t]$ to remove transient behavior. These macroscopic variables are the complex order parameter $R=|Z|$ given by \eqref{eq:Z} and the weighted order parameter $S=|W|$ given by \eqref{eq:W}.
Inspecting the phase diagram there appears to be a transition between \LD\ and \LL\ states at a a critical $p=p_c$. 
An incoherent state with $R=S=0$ is not observed. In the following, we attempt to explain these observations by analyzing the dynamics described by \eqref{eq:r1}-\eqref{eq:delta}.

\subsection{Reduced dynamical equations}
We explain the observed behavior by studying the reduced equations~\eqref{eq:z_dynamics} describing the dynamics of the local order parameters \eqref{eq:z_uncorrelated} valid for the continuum limit with the $M=2$ populations present in the correlated model. 
Observing the phase shift invariance of the system, we can further reduce one dimension by introducing the phase difference $\delta \equiv \zangle_2 - \zangle_1$, resulting in the system of differential equations given by
\begin{subequations}
\begin{align}\label{eq:r1}
    \dot{r}_1 &= \frac{1-{r^2_1}}{2} \Bigg(p r_1 -q r_2 \cos\delta \Bigg), \\ 
    \label{eq:r2}
    \dot{r}_2 &= \frac{1-{r^2_2}}{2} \Bigg(p r_1 \cos\delta-q r_2\Bigg), \\ 
    \label{eq:delta}
    \dot{\delta} &= \gamma-\Bigg(\frac{1+{r^2_2}}{2r_2} p r_1 - \frac{1+{r^2_1}}{2r_1}q r_2 \Bigg)
    \sin\delta.
\end{align}
\end{subequations}

\subsection{Equilibrium states\label{sec:corr_equilibria}}
\subsubsection{Incoherent (\INC) state}
The incoherent state (\INC) is defined by $ R = S = 0 $. Recalling Eq.~\eqref{eq:W_correlated} and \eqref{eq:ZZ_correlated}, we immediately recognize that these conditions result from letting $ r_1 = r_2 = 0$ or $z_1=z_2=0$. Eqs.~\eqref{eq:r1}-\eqref{eq:delta} are given in polar coordinates and are hence singular in this point; we therefore instead inspect Eqs.~\eqref{eq:z_dynamics} for $M=2$ in complex coordinates and note that this incoherent state exists for any parameter choice.  
The associated Jacobian 
\begin{align}
    J_\text{INC}&=
    \left(
    \begin{array}{cccc}
    p/2 & \gamma  q & -q/2 & 0 \\
    -q\gamma  & p/2 & 0 & -q/2 \\
    p/2 & 0 & -q/2 & -\gamma p  \\
    0 & p/2 & \gamma  p & -q/2 \\
    \end{array}
    \right)
\end{align}
has two pairs of complex conjugated eigenvalues, 
\begin{subequations}
\begin{align}\nonumber
    \lambda_{1,2}&=
    \frac{1}{4} \Big(
    -1 
    + 2 p
    + 2i\gamma|1-2p|\\
    &\pm\sqrt{(1-2p)^2-4\gamma^2 + \text{sgn}(1-2p) 4i\gamma}
    \Big),\\\nonumber
    \lambda_{3,4}&=
    \frac{1}{4} \Big(
    -1 
    + 2 p
    - 2i\gamma|1-2p|\\
    &\pm\sqrt{(1-2p)^2-4\gamma^2 - \text{sgn}(1-2p) 4i\gamma}
    \Big).\
\end{align}
\end{subequations}
Inspecting these eigenvalues numerically reveals that INC is unstable for almost all parameter choices:
the eigenvalues are complex-valued with $ \Re{\lambda_k} > 0$ for $ 0 < p \leq 1 $ and $ \gamma>0 $;  exceptional cases occur for two cases, namely, for $ p = 0 $, where INC is stable; or for $ p < \half $ and $ \gamma = 0 $, where INC is neutrally stable.

\subsubsection{Lock-Lock (\LL) state}
Next we examine the {\it{Lock-Lock}} (\LL) state where oscillators in both subpopulations $\Subpop_1$ and $\Subpop_2$ are phase locked, i.e., $r^*_1=r^*_2=1$.
\begin{figure}
 \centering
 \includegraphics[width=0.85\columnwidth]{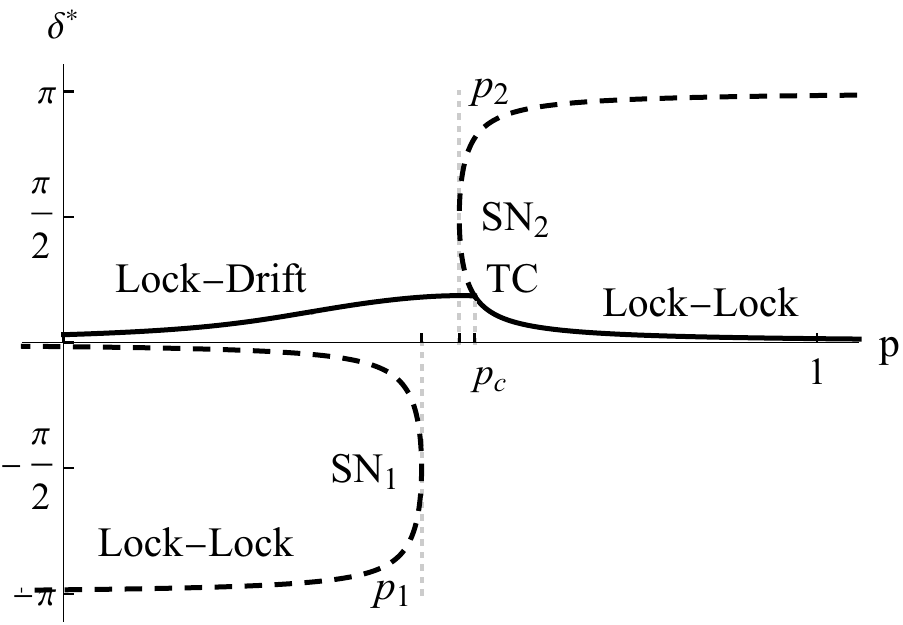}\\\vspace{1em}
 \includegraphics[width=0.85\columnwidth]{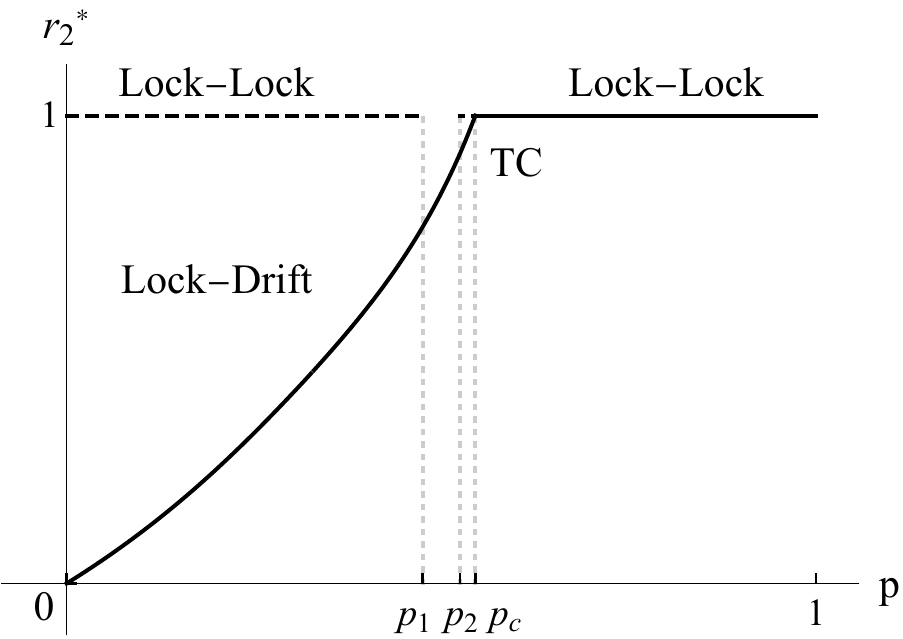}
 \caption{Bifurcation diagram for Lock-Drift and Lock-Lock states in the correlated model for $\gamma=0.05$, with stable (solid) and unstable (dashed) branches. Saddle node bifurcations (SN$_1$ and SN$_2$) occur at $p_1$ and $p_2$, and a transcritical bifurcation at $p=p_c$ (see text).
 \label{fig:FixedPoints_LD_delta}
 }
\end{figure}
These conditions immediately satisfy the fixed point conditions for \eqref{eq:r1} and \eqref{eq:r2} by definition; Eq.~\eqref{eq:delta} yields the fixed point condition
\begin{eqnarray}
    \label{eq:FixedPoints_LL_delta}
    && \sin\delta^* = \frac{\gamma}{p-q}=\frac{\gamma}{2p-1},
\end{eqnarray}
with the explicit solutions
\begin{align}\label{eq:LLdelta}
 \delta^*_{{+,-}} &= 
 \left\{
 \begin{array}{l}\smallskip
    \arcsin{\left(\dfrac{\gamma}{2p-1}\right)}+2\pi  k\\
    \pi-\arcsin{\left(\dfrac{\gamma}{2p-1}\right)}+2\pi k \\
 \end{array}
 \right.,
 \
\end{align}
where $k\in \mathbb{Z}$ {and subscripts `+' and `-' label the two solution branches for~\eqref{eq:delta}.}
These two solution branches are born in saddle-node bifurcations (SN$_1$ and SN$_2$) located at $p_1:=|\gamma-1|/2$ and $p_2:=|\gamma+1|/2$, respectively. Thus, the existence of these branches is limited to $0\leq p<p_1$ and $p_2<p\leq 1$ (see Fig.~\ref{fig:FixedPoints_LD_delta}), and consequentially to $\gamma \leq 1$ since $ p_2 \leq 1 $.

The Jacobian of \eqref{eq:r1}-\eqref{eq:delta} for the \LL\ state can be expressed as 
\begin{align}
    J_\text{LL}&=
    \left(
    \begin{array}{ccc}
    -p + (1-p)\cos{\delta^*} & 0 & 0 \\
    0 & 1-p-p \cos{\delta^*} & 0  \\
    -p\sin{\delta^*} & (1-p)\sin{\delta^*} &  (1-2p)\cos{\delta^*} \
    \end{array}
    \right)
\end{align}
where $\delta^*$ is given by ~\eqref{eq:LLdelta}.
The Jacobian is tri-diagonal and we readily obtain the eigenvalues for \LL\ by substituting the two solution branches for $\delta^*$ and eliminating $q$,  
\begin{subequations}
\begin{align}
    \lambda_1^{+,-} &= 
    \pm (2 p - 1) \sqrt{1-\frac{\gamma ^2}{(1-2 p)^2}},\\
    \lambda_2^{+,-} &= 
    1 \pm p \left(\sqrt{1-\frac{\gamma ^2}{(1-2 p)^2}}-1\right),\\
    \lambda_3^{+,-} &= 
    - p 
    \pm (p-1)
    \sqrt{1-\frac{\gamma ^2}{(1-2 p)^2}}.\
\end{align}
\end{subequations}
Plotting real and imaginary of these eigenvalues reveals that only the second branch is linearly stable for $p_c<p\leq 1$, where $p_c$ denotes the critical point $p_c$  where \LL\ state loses stability and connects to the \LD\ state.

\subsubsection{Lock-Drift (\LD) state}
We examine the {\it{Lock-Drift}} (\LD) state, where oscillators in the first subpopulation ($\Subpop_1$) with $\omega^{(1)}=-q\gamma$ and $\xi^{(1)}=+1$ show perfect synchronization, $ r_1 = 1 $, while oscillators in the second subpopulation ($\Subpop_2$) with $\omega^{(2)}=p\gamma$ and $\xi^{(2)}=-1$ are drifting incoherently with $r_2 < 1 $. Thus, the \LD\ state may appear like the symmetry breaking ``chimera state'' known from previous studies~\cite{Abrams2008,Martens2016} in the sense that one subpopulation of the oscillators displays perfect synchronization, but the other does not; however, the \LD\ state occurring in Eqs.~\eqref{eq:model} has a different origin, since it arises due to the correlation of the two disorders, $\omega_i$ and $\xi_i$; moreover, unlike the chimera state, the \LD\ state does not have a symmetric counterpart corresponding to a \DL\ state (see Sec.~\ref{sec:corr_absence_L-D_state}).

Fixed point conditions for Eqs.~\eqref{eq:r1}-\eqref{eq:delta} are satisfied for the \LD\    state with $r_1 = 1$ if in addition we demand stationary $\delta=\delta^*$ and $r^*_{2}\neq 1$, i.e.,
\begin{subequations}
\begin{align}\label{eq:r1r2_smallp}
    r_2^*    &= \frac{p}{q}\cos{\delta^*},\\
    \gamma &= \sin{\delta^*}\left(\frac{1+r_2^2}{2r_2}p - q r_2^* \right).\
\end{align}
\end{subequations}
While it is possible to eliminate $r_2$ such as to obtain an equation containing $\delta$ as the only variable, we instead eliminate $\delta^*$ by using $1=\cos^2{\delta^*}+\sin^2{\delta^*}$ and solving the two conditions above for 
\begin{subequations}
\begin{align}\label{eq:LDcosdelta}
    \cos{\delta^*}&=\frac{q}{p}r_2^*,\\\label{eq:LDsindelta}
    \sin{\delta^*}&=\gamma\,\frac{2r_2^*}{(1+(r_2^*)^2)p-2q(r_2^*)^2},
\end{align}
\end{subequations}
resulting in 
\begin{align}
    1&=T \left(\frac{(p-1)^2}{p^2}+\frac{4 \gamma ^2}{((3 p-2) T+p)^2}\right)\
\end{align}
where $ T:=(r_2^*)^2 \geq 0 $. This cubic polynomial can be solved for $ r_2^*=+\sqrt{T} $ using computer assisted algebra, resulting in one real and two complex conjugated roots --- too unwieldy to display here. Finally, we obtain from \eqref{eq:r1r2_smallp} the fixed point solution shown in Fig.~\ref{fig:FixedPoints_LD_delta},
\begin{align}
    \delta^* &= \arccos{\left(\frac{q}{p}r_2^*\right)},\
\end{align}
where only the positive branch in Eq.~\eqref{eq:r1r2_smallp} is a valid solution since $ p,q,r_2\geq 0$ must be non-negative. The \LD\ state exists for $ 0 \leq p < p_c$, where $ p = p_c $ defines the transition from \LD\ to \LL\ state computed in Sec.~\ref{sec:corr_stability_diagram} further below. Numerically plotting the eigenvalues of this branch reveals that they are real and negative for all $ 0 \leq \gamma\leq 1 $ with $ 0 \leq  p < p_c $. While fixed points do exist for $ p > p_c $, they are not physically meaningful since they have $ r_2 > 1 $ (we therefore do not show this branch in Fig.~\ref{fig:FixedPoints_LD_delta}). However, their eigenvalues have positive real parts, thus prompting a transcritical bifurcation, denoted TC, at $ p = p_c $.  Furthermore, we observe that $r_2$ is monotonically increasing for $p<p_c$, but monotonically decreasing for $p>p_c$; as a consequence, the peak value of the relative phase between the two subpopulations is reached at $p=p_c$ where $r_2=1$ so that $\delta=\arccos{(q/p)}$. These results are summarized in Fig.~\ref{fig:stability_diagram}.

\subsubsection{Absence of Drift-Lock state}\label{sec:corr_absence_L-D_state}
The \emph{``Drift-Lock (D-L)''} state with $r_1=0$ and $r_2>0$ does not emerge in the system. This is easily seen as follows. The oscillators in the first subpopulation $\Subpop_1$ with positive coupling strength $\xi^{(1)}=+1$ tend to minimize their phase difference, thus resulting in 
phase-locking behavior.  On the other hand, oscillators in the second subpopulation $\Subpop_2$ with negative coupling strength $\xi^{(2)}=-1$ tend to 
maximize the phase differences, thus displaying drifting behavior.  

\subsubsection{Stability diagram }\label{sec:corr_stability_diagram}
\begin{figure}[htp!]
 \centering
 \includegraphics[width=.85\columnwidth]{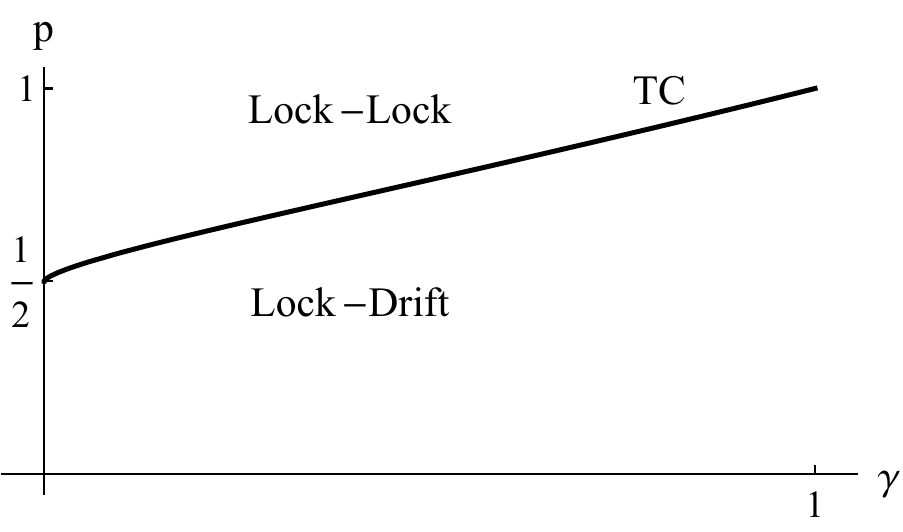}
 \caption{\label{fig:stability_diagram}Stability diagram for the correlated model with infinite oscillators. The stability boundary $(p_c,\gamma_c)$ between lock-lock (\LL) and lock-drift states (\LD) given by \eqref{eq:pcrit_gammacrit} corresponds to a transcritical bifurcation (see text and Fig.~\ref{fig:FixedPoints_LD_delta}).}
\end{figure}

We establish a stability diagram for the three states discussed above: incoherence (\INC), lock-lock (\LL), lock-drift (\LD). 
We have already shown that {\INC\ can} only be (neutrally) stable for $p=0$ (or $\gamma=0$ with $p<1/2$); we are left to determining the transition point between lock-lock and lock-drift states, i.e., the critical value $p_c$ at which the transition between stable \LD\ and \LL\ states occurs.
To do this, we consider the fixed point condition for the \LL\ state, Eq.~\eqref{eq:FixedPoints_LL_delta}, to be considered in the limit from above where $p \rightarrow p_c^+$ and $\delta^*\rightarrow \delta_c$ ; and the fixed point condition for the lock-drift state, Eq.~\eqref{eq:LDcosdelta}, in the limit from below where $p\rightarrow p_c^-$ and $r_2^*\rightarrow 1^-$. At this point, we have 
\begin{align*}
    \cos{\delta_c} &= \frac{q_c}{p_c} \quad \text{and} \quad
    \sin{\delta_c} = \frac{\gamma_c}{2p_c-1},\
\end{align*}
for the \LL\ and \LD\ states, respectively; both fixed point conditions satisfy $1=\cos{(\delta_c)}^2+\sin{(\delta_c)}^2$ simultaneously, so that 
\begin{align}
    1 &= \left(\frac{1-p_c}{p_c}\right)^2 + \left(\frac{\gamma}{2p_c-1}\right)^2,\
\end{align}
which is equivalent to
\begin{align}
    8 {p_c}^3 - (12+\gamma_c^2) {p_c}^2 + 6p_c -1 =0,
    \label{eq:pcrit}
\end{align}
provided that $p_c \neq 0, p_c \neq \half$.
Since $\gamma>0$, we may infer the relationship
\begin{align}\label{eq:pcrit_gammacrit}
    \gamma_c &= \frac{(2p_c-1)^{3/2}}{p_c},\
\end{align}
which produces the stability diagram in Fig.~\ref{fig:stability_diagram}.

We find that  $p_c$ monotonically increases as $\gamma=\gamma_c$ increases, which is reasonable in the sense that a higher value of $p$ is required to make the oscillators synchronized for a wider distribution with increasing value of $\gamma$. Since values $p>1$ are not meaningful so that $\gamma=1$ constitutes an absolute limit for the existence of the \LD\ state.

\subsubsection{Global order parameters and traveling waves}

We investigate the behavior of order parameters $W$ and $Z$ for the two stable equilibria found, \LD\ and \LL. In the limit of infinite oscillators, where $ N_1/N \rightarrow p$ and $ N_2/N \rightarrow q$, the complex order parameter~\eqref{eq:W_correlated} becomes
\begin{align}
    Z  = R e^{i\Psi} &=  p z_1 + q z_2 = pr_1 e^{i\zangle_1} + qr_2 e^{i\zangle_2},
\end{align}
which has magnitude 
\begin{align}
    R &= \sqrt{p^2 r_1^2 + 2pq r_1 r_2 \cos{\delta}  + q^2 r_2^2};
    \label{eq:R}
\end{align}
similarly, the weighted order parameter~\eqref{eq:W_uncorrelated} is
\begin{align}
    W &= S e^{i\Delta} 
      &= p z_1 - q z_2 = pr_1 e^{i\zangle_1} - qr_2 e^{i\zangle_2},
\end{align}
with magnitude 
\begin{align}
    S &= \sqrt{p^2 r_1^2 - 2pq r_1 r_2 \cos{\delta} + q^2 r_2^2 }.
\label{eq:S}
\end{align}
Furthermore, we may determine the mean-field frequency or ``wave speed'' of the collective state (see App.~\ref{sec:AppendixB} for a derivation),
\begin{align} 
    \Omega &:= \left|\frac{\d}{\d t}{\Delta}\right| = \frac{\d}{\d t}{\arg{(W)}} = \frac{1}{S} \sqrt{|\dot{W}|^2-\dot{S}^2}, 
    \label{eq:Omega}
\end{align} 
where 
\begin{align}\label{eq:dWdt}
    \begin{split}
    |\dot{W}|^2 &= p^2(\dot{r_1}^2+r_1^2\dot\zangle_1^2) +q^2(\dot r_2^2+r_2^2\dot\zangle_2^2) \\
    &+ 2pq [(\dot r_1 r_2 \dot\zangle_2 - r_1\dot r_2 \dot\zangle_1)\sin\delta 
    - ({\dot r_1}{\dot r_2}+ r_1 r_2 {\dot\zangle_1}{\dot\zangle_2})\cos\delta] \
    \end{split}
\end{align}
and 
\begin{align}
    \begin{split}
    \dot{S} &= \frac{1}{S} \Bigg( p^2 r_1 \dot r_1 + q^2 r_2 {\dot r_2} \\
        &- pq \left((\dot r_1 r_2+r_1{\dot r_2})\cos\delta -r_1 r_2 {\dot{\delta}}\sin\delta \right) \Bigg), \label{eq:dSdt}
    \end{split}
\end{align}

with $\delta=\zangle_2-\zangle_1$. 
Evaluating $R$, $S$ and $\Omega$ at the equilibria corresponding to \LD\ and \LL\ states, we are able to plot the behavior of $R$ and $S$ as a function of $p$ for $\gamma=0.05$ as shown in Fig.~\ref{fig:R_S}. 

It should be clear that the nature of the \LD\ state occurring for $p<p_c$ implies that $R<1$; however, note that the \LL\ state occurring for $p>p_c$ does not necessarily imply perfect synchronization for the complete system in the sense that $ R = 1 $, since the locked oscillators of the two subpopulations may assume non-identical mean-field phases, ($\delta = \varphi_2-\varphi_1 \neq 0$), which results in $R < 1$. Inspecting Fig.~\ref{fig:FixedPoints_LD_delta} we recognize that $R=1$ is only achieved for $p=1$ where $\delta=0$. Indeed, evaluating the order parameter for the \LL\ state, the asymptotic behavior  for $p$ close to 1 is $R  \sim 1 - \left(1-\sqrt{1-\gamma ^2}\right) (1-p) + \mathcal{O}\left((1-p)^2\right)$.
Note that $R=|Z|=0$ is only possible if $|z_1|=|z_2|=0$ as long as $p>0,q>0$; however, we found that such an \INC\ is (almost always) unstable. As a consequence, we can also rule out the case where $S=|W|=0$ or $S>0$ with $R=0$. 
Furthermore we note that the nonzero wave speed, $\Omega \neq 0$, seen in {Fig.~\ref{fig:R_S}} implies the presence of the traveling wave studied in Ref.~\cite{Hong2016}: {thus,} we confirm that the asymmetry in the correlated disorder induces the motion of a traveling wave{, rather than being induced by other type of heterogeneity}. {While the wave speed could be set to zero by an appropriate choice of reference frame, we note that the wave speed $\Omega$ differs from the system's mean natural frequency.}

\section{Analysis for uncorrelated disorder}\label{sec:uncorrelated_model_analysis}

\subsection{Numerical simulations}\label{sec:uncorrelated_num_sim}
\begin{figure}[!htpb]
    \centering
    \includegraphics[width=0.95\linewidth]{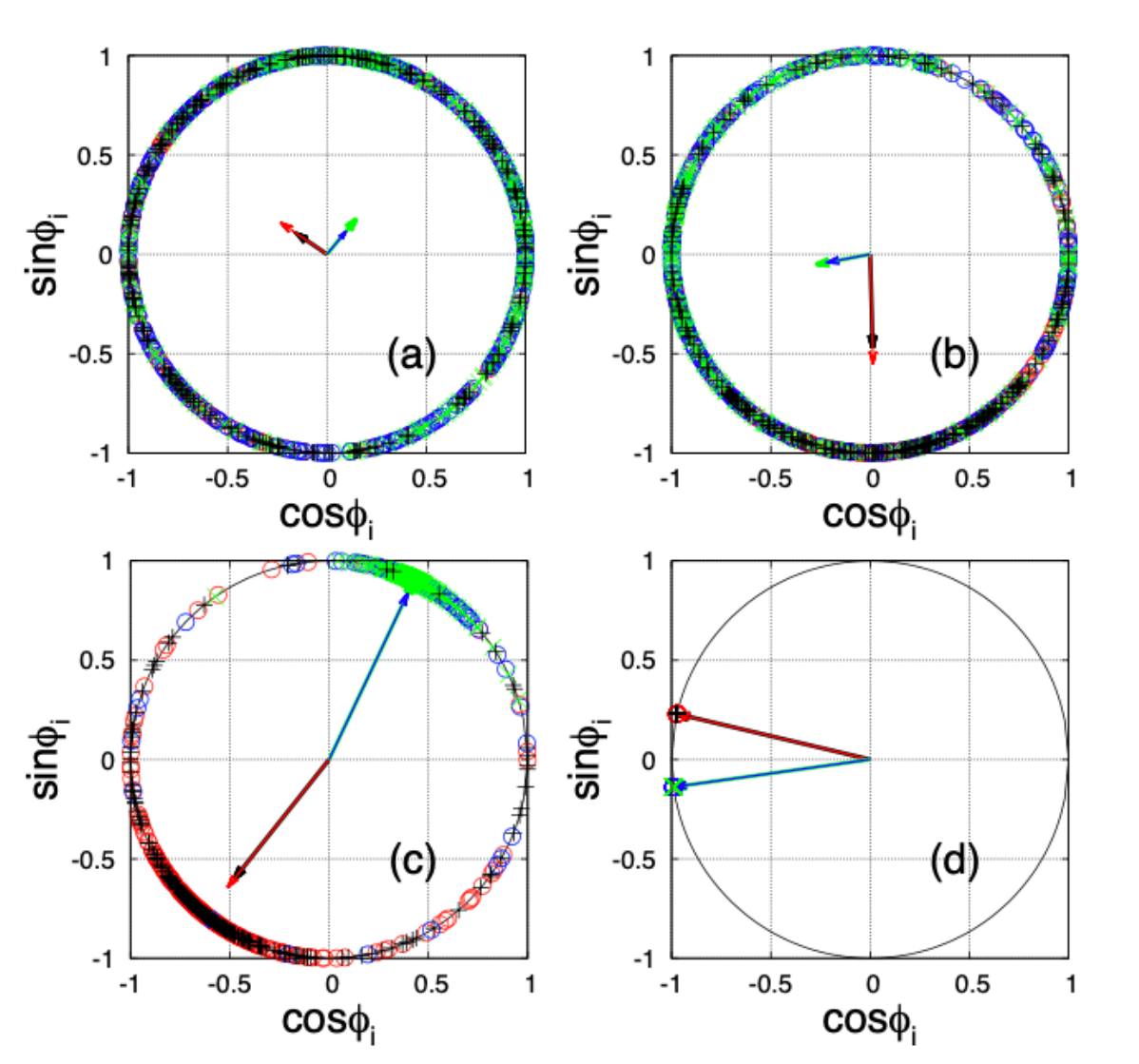}
    \caption{\label{fig:phi_uncorrelated} (Color Online) Snapshot of the phases $\phi_i$ are shown in the plane of $(\cos\phi_i, \sin\phi_i)$ for $N=1000$ oscillators for the uncorrelated model with $\gamma=0.05$. 
    \INC\ states are observed for $p=0.4$ (a), $p=0.45$ (b), $p=0.49$ (c); the \LLLL\ state is observed for $p=0.55 (>p_c)$ in panel (d), 
    and $p_c$=1/2 is the critical point where the \INC\ state loses stability.      The symbols $\textcolor{red}{\circ},\textcolor{blue}{\circ},\textcolor{green}{\times},+$  represent phases of the oscillators belonging to $\Subpop_1$, $\Subpop_2$,  $\Subpop_3$, $\Subpop_4$, respectively, and red/blue/green/black arrows show local order parameters $z_1$, $z_2$, $z_3$, $z_4$, respectively.
    }
\end{figure}

For the uncorrelated model, we first performed numerical simulations of Eq.~\eqref{eq:model} using a fourth-order Runge-Kutta (RK4) integration scheme with identical parameters as listed in Sec.~\ref{sec:correlated_num_sim} for the correlated model. 
Snapshots of asymptotic states are shown in Fig.~\ref{fig:phi_uncorrelated} for several values of $p$. We first observed that, for all reported values, oscillators residing in the subpopulations $\Subpop_1$ and $\Subpop_4$,  and in the subpopulations $\Subpop_2$ and $\Subpop_3$, respectively, are phase-locked. We found that the system may exhibit at least three states:
\begin{itemize}
 \item [i)] The \emph{Incoherent state} (\INC) where all subpopulations are desynchronized, i.e, $r_1\approx r_2\approx r_3\approx r_4\approx 0$\footnote{Deviations in numerical simulations are due to finite size effects and critical slowing down near $p=p_c$}, see panels a), b) and c).
 
 \item [ii)] The \emph{Breathing Lock-Lock-Lock-Lock state} (Breathing \LLLL)  where oscillators in each subpopulations are phase-locked, $r_1=r_2=r_3=r_4=1$, but where the two mutually phase-locked subpopulation pairs $(\Subpop_1,\Subpop_4)$ and $(\Subpop_2,\Subpop_3)$ drift apart, i.e., their phase difference $\delta(t):=\phi_2(t)-\phi_1(t)$ increases with time. 
 
 \item [iii)] The \emph{Stable Lock-Lock-Lock-Lock state} (Stable \LLLL) where oscillators in each subpopulations are phase-locked, $r_1=r_2=r_3=r_4=1$, and the phase-locked subpopulation pairs $(\Subpop_1,\Subpop_4)$ are frequency locked, i.e. their phase difference remains constant in time,  $\dot{\delta}=0$, see panel d). 
\end{itemize}

\begin{figure}[!htp]
    \includegraphics[width=0.95\linewidth]{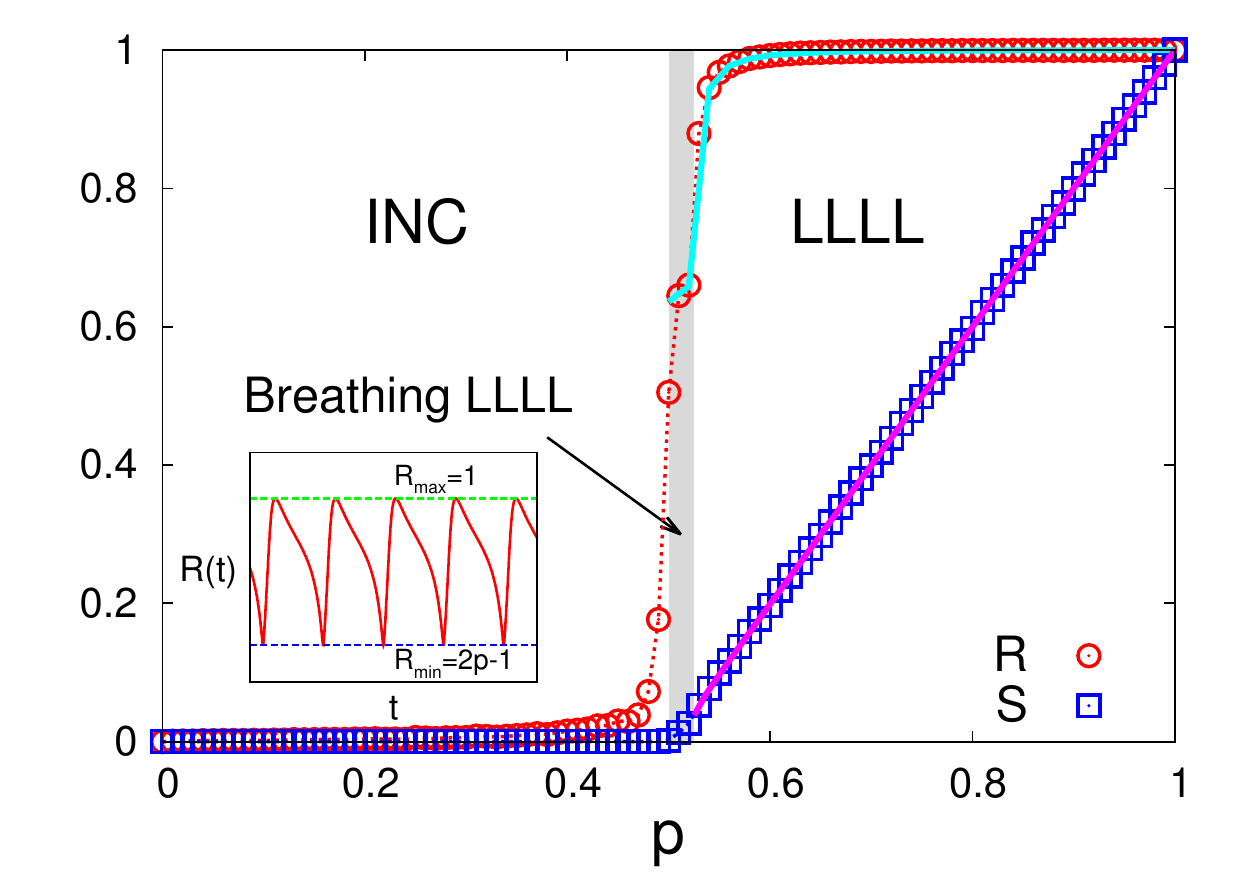}
    \caption{\label{fig:simulation}(Color Online) Phase diagram for the uncorrelated model obtained via numerical simulation of Eqs.~\eqref{eq:model}. Equilibrium values for the order parameters $R$ (red circles) and $S$ (blue squares) are shown as a function of $p$ for $\gamma=0.05$ after transient behavior has vanished (see text).  The system size is $N=10^5$ and the data represent values averaged over 10 sample simulations with different initial conditions $\{\phi_i(0)\}$.
    The incoherent state \INC\ with $R=S=0$ is observed for $p<p_c$, while the coherent state \LLLL\ with $R>0$ and $S>0$ emerges for $p>p_c$, where $p_c = 1/2$. Theoretical predictions for $R$ (magenta) given by Eq.~\eqref{eq:R_uncorrelated_4D_OA} and for $S$ (cyan) given by Eqs.~\eqref{eq:selfcons_S_sol} and \eqref{eq:S_uncorrelated_4D_OA}, valid for $p>p_c$, match the results obtained from numerical simulations very well. The inset shows the periodic behavior of $R(t)$ in time for $p=0.52$ (Breathing \LLLL).
    \label{fig:RS_nocorr}}
\end{figure}

We also measured the asymptotic behavior for the order parameters, $R$ and $S$, averaged over the time window $[M_t/2,M_t]$ to quantify the collective synchronization level of the system, while varying the probability $p$. 
Fig.~\ref{fig:simulation} shows the resulting time asymptotic behavior of $R$ and $S$ while varying $p$ for $\gamma=0.05$ fixed.

The incoherent (INC) state ($R=S=0$) appears to exist only for $0 < p < p_c$, while the coherent {\LLLL\  states} exist for $p>p_c$. {However, note that the Ott/Antonsen Eqs.~\eqref{eq:zeq_uncorrelated_1}-\eqref{eq:zeq_uncorrelated_4} reveal neutral stability of the incoherent state, as we discuss further below (Sec.~\ref{sec:Transverse_Stability_UncorrModel}). Moreover, random initial conditions for the local order parameters $z_\sigma(0)$ evolve to arbitrary asymptotic order parameter values with $r_\sigma(t)>0$. Thus, we expect that initial phases  deviating more strongly from $R=S=0$ in \eqref{eq:model} also asymptotically evolve to values with $R>0,S>0$, incongruent with Incoherence.
}

The critical value for this transition, $p_c$, may be deduced from a simple argument: we expect that the coherent state with $R>0$ only exists for coupling strengths with positive mean given by 
\begin{equation}
    \langle \xi \rangle = p \cdot 1 + q \cdot (-1) = 2p-1>0.
\end{equation}
Since we cannot expect that the coherent state emerges for repulsive coupling, $\langle \xi \rangle<0$, we obtain the critical value $p_c=1/2$.
Note that in the present study we chose $\xi_j=1$ for $j\in\Subpop_1$ and $ \xi_j = -1 $ for $j\in\Subpop_2$ without loss of generality. 
We may instead assign general asymmetrically balanced values {($\xi_+\neq\xi_-$)}, $ \xi = \xi_{+} > 0 $ with probability $ p $ and  
$ \xi = \xi_{-} < 0 $ with probability $ 1-p $. Then we have
\begin{eqnarray}
    \langle \xi \rangle &=& p\,\xi_+ + (1-p) \,\xi_- 
    = (p+(p-1)Q)\,\xi_+
\end{eqnarray}
where we define $ Q:= -\xi_{-}/\xi_{+}>0$. Again, the coherent state exists for $\langle\xi\rangle > 0$ only, thus determining a critical value given by $p_c=Q/(1+Q)$.
Applying this to the present case with $\xi_j=\pm 1$ results  in $Q=1$ which yields
our previous critical value of $p_c=1/2$, as expected. This value agrees well with our numerical simulations, see Fig.~\ref{fig:simulation}. 

In the following we explain the observed behavior using the dimensionally reduced equations derived in Sec.~\ref{sec:OttAntonsen} and a self-consistency argument.

\subsection{Reduced dynamical equations}
We explain the observed behavior by studying Eqs.~\eqref{eq:z_dynamics} describing the dynamics for the local order parameters in \eqref{eq:z_uncorrelated} valid for the continuum limit with the $M=4$ populations present in the uncorrelated model, given by
\begin{subequations}
\begin{align}\label{eq:zeq_uncorrelated_1}
    \dot{z}_1 &= +iq\gamma z_1 + \frac{1}{2}\left(W - \bar{W}{z^2_1}\right), \\\label{eq:zeq_uncorrelated_2}
    \dot{z}_2 &= -ip\gamma z_2 + \frac{1}{2}\left(W - \bar{W}{z^2_2}\right), \\\label{eq:zeq_uncorrelated_3}
    \dot{z}_3 &= -ip\gamma z_3 + \frac{1}{2}\left(W - \bar{W}{z^2_3}\right), \\\label{eq:zeq_uncorrelated_4}
    \dot{z}_4 &= +iq\gamma z_4 + \frac{1}{2}\left(W - \bar{W}{z^2_4}\right),
\end{align}
\end{subequations}
where the weighted order parameter is given by~\eqref{eq:W_uncorrelated},
\begin{align*}
    W&=p^2 z_1 - pqz_4 + pq z_3 - q^2z_2.\
\end{align*}

We note that Eqs.~\eqref{eq:zeq_uncorrelated_1} and \eqref{eq:zeq_uncorrelated_4} for subpopulations $\Subpop_1$ and $\Subpop_4$, and  Eqs.~\eqref{eq:zeq_uncorrelated_2} and \eqref{eq:zeq_uncorrelated_3}  for subpopulations $\Subpop_2$ and $\Subpop_3$, have identical structure. Furthermore, numerical simulations (Sec.~\ref{sec:uncorrelated_num_sim}) 
revealed asymptotic behavior for the \LLLL\ states, i.e., $|z_1(t)-z_4(t)|\rightarrow 0$ and $|z_2(t)-z_3(t)|\rightarrow 0$ as $t\rightarrow \infty$. This observation suggests the existence of a stable symmetric invariant subspace \SS\ defined by $z_1(t)=z_4(t)$ and $z_2(t)=z_3(t)$ for all $t$. We therefore first examined the dynamics confined to that subspace. Eqs.~\eqref{eq:zeq_uncorrelated_1} and \eqref{eq:zeq_uncorrelated_2} govern this dynamics. Introducing polar coordinates $z_\sigma = r_\sigma e^{i\varphi_\sigma}$ and defining $\delta := \varphi_2-\varphi_1$,  we have
\begin{subequations}
\begin{align}\label{eq:OA_uncorrelated_polar}
    \dot{r}_1 &= \frac{1-r_1^2}{2}(p-q)(p r_1 + q r_2 \cos{\delta}),\\
    \dot{r}_2 &= \frac{1-r_2^2}{2}(p-q)(p r_1\cos{\delta} + q r_2), \\
    \dot{\delta} &= \gamma
    -\half(p-q)
    \left(
    \frac{1+r_2^2}{r_2} p r_1
    +
    \frac{1+r_1^2}{r_1} q r_2
    \right)\sin{\delta}.\
\end{align}
\end{subequations}

\subsection{Equilibrium states}
\subsubsection{Incoherent (INC) state}
The incoherent state is defined by $z_1=z_2=z_3=z_4=0$. The Jacobian for Eqs.~\eqref{eq:zeq_uncorrelated_1} and 
\eqref{eq:zeq_uncorrelated_2} describing the dynamics in $z_1$ and $z_2$ on the symmetric subspace \SS, defined by $z_1(t)=z_4(t)$, $z_2(t)=z_3(t)$, expressed in Cartesian coordinates is
\begin{align}
    J_\text{INC}&=
    \left(
    \begin{array}{cccc}
    p/2 & \gamma  q & -q/2 & 0 \\
    -\gamma q  & p/2 & 0 & -q/2 \\
    p/2 & 0 & -q/2 & -\gamma p \\
    0 & p/2 & \gamma  p & -q/2 \\
    \end{array}
    \right)
    \end{align}
has two pairs of complex conjugated eigenvalues, 
\begin{subequations}
\begin{align}\nonumber
    \lambda_{1,2}&=
    \frac{1}{4} \Bigg(
    2 p-1
    +2 i \gamma  | 1-2 p|\\
    &\pm\sqrt{4 i \gamma  (2 p-1) | 1-2 p| -4 \gamma ^2+(1-2 p)^2}
    \Bigg)\\\nonumber
    \lambda_{3,4}&=
    \frac{1}{4} \Bigg(
    2 p-1
    -2 i \gamma  | 1-2 p|\\
    &\pm\sqrt{4 i \gamma  (2 p-1) | 1-2 p| -4 \gamma ^2+(1-2 p)^2}
    \Bigg)
\end{align}
\end{subequations}
Inspecting $\Re{\lambda_k}$ numerically for $k=1,2,3,4$ we immediately see that \INC\ is stable on the symmetric subspace \SS\ only when $p<p_c=1/2$; otherwise, it is unstable.

\subsubsection{Stable and breathing \LLLL\ states}
\begin{figure}[htp!]
 \centering
 \includegraphics[width=0.85\columnwidth]{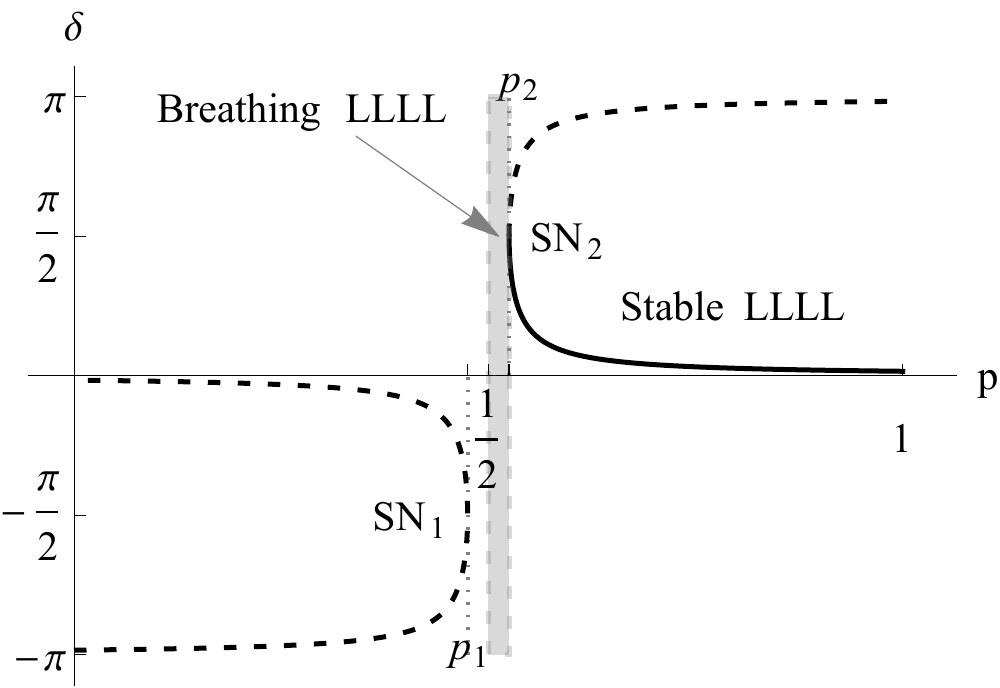}\\
  \includegraphics[width=0.85\columnwidth]{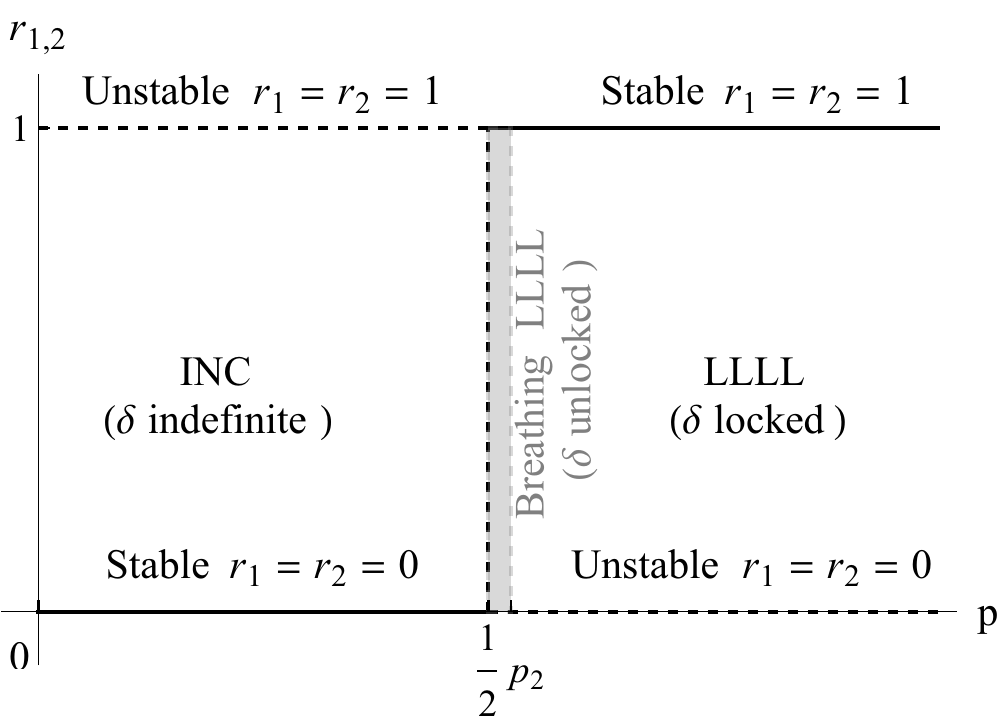}\\
 \caption{Bifurcation diagram for uncorrelated disorder ($\gamma=0.05$). 
 Stable and unstable branches of \INC\ and \LLLL\ states are indicated as solid and dashed curves, respectively (Stability relates to the symmetric subspace \SS\ given by Eqs.~\eqref{eq:OA_uncorrelated_polar}). Breathing \LLLL\ states corresponding to limit cycles on the subspace $r_1=r_2=1$ are anihilated in the saddle-node bifurcation SN$_2$.
 \label{fig:Bifurcation_diagram_uncorrelated_model}
 }
\end{figure}

Locked states (\LLLL) satisfy $r_1=r_2=r_3=r_4=1$.
This also defines an invariant subspace (on the symmetric subspace \SS) since the \LLLL\ state implies
$ \dot{r}_1=\dot{r}_2 = 0 $. In the folowing, we consider the dynamics and stability of \LLLL\ states on \SS\ as given by Eqs.~\eqref{eq:OA_uncorrelated_polar}. 
Stationarity of the \LLLL\ state requires the additional condition $\dot{\delta}=0$, which implies the stationary phase difference 
\begin{align}\label{eq:deltastar}
    \sin{\delta^*} = \frac{\gamma}{2p-1}.
\end{align}
We denote an equilibrium with $(r_1,r_2,\delta)=(1,1,\delta^*)$ as a \emph{Stable  \LLLL} state. Eq.~\eqref{eq:deltastar} informs us that stable \LLLL\ states are born in saddle-node bifurcations SN$_1$ and SN$_2$ at $p_1=|\gamma-1|/2$ and $p_2=|\gamma+1|/2$ and are constrained to the intervals $ 0\leq p \leq  p_1 $ and $ p_2 \leq p \leq 1 $ (see Fig.~\ref{fig:Bifurcation_diagram_uncorrelated_model}). 

To examine stability, consider the eigenvalues of the Jacobian for \LLLL,
\begin{subequations}
\begin{align}
    \lambda_1 &= (1-2p) \cos{\delta^*},\\
    \lambda_2 &= (1-2p) (1 + p (1-\cos{\delta^*})),\\
    \lambda_3 &= (1-2p) (p + (1-p) \cos{\delta^*}).\
\end{align}
\end{subequations}
We first note that all eigenvalues flip sign at $p=p_c=1/2$. It  therefore suffices to consider eigenvalues restricted to the interval $1/2 \leq p\leq 1$ where they share the common factor $(1-2p)<0$.
For $p\in [p_2,1]$, the lower branch $\delta^*\in[0,\pi/2]$ has $\Re{\lambda_1(\delta^*)}<0$, whereas the upper branch with $\delta^*\in[\pi/2,\pi]$ has $\Re{\lambda_1(\delta^*)}>0$.
Since $0\leq \delta^* \leq \pi$ for $p\in[p_2,1]$ we have $0\leq \cos{\delta^*}\leq 1$ and $\lambda_2$, $\lambda_3$ are real-valued. Minimizing and maximizing these two eigenvalues, we find  that $(1-2p) (1+p)<\lambda_2<(1-2p)<0$ and $(1-2p)p<\lambda_3<(1-2p)$.
As already mentioned, the signs of all eigenvalues are reversed for $p<1/2$. 
Therefore, the { \LLLL} state $(1,1,\delta^*)$ is stable for $p\in[p_2,1]$, as shown in Fig.~\ref{fig:Bifurcation_diagram_uncorrelated_model}. 

For $p_1\leq p \leq p_2$, the phase difference $\delta(t)$ is unlocked and evolves according to
\begin{align}
\dot{\delta} &= \gamma
+(2p-1)\sin{\delta},\
\end{align}
We denote the resulting limit cycle,  confined to the invariant suspace $r_1=r_2=1$, as the {\it Breathing} \LLLL\ state. 

Furthermore, for $p=1/2$ we have $\dot{r}_1 = \dot{r}_2 \equiv 0$ (with $\delta = \gamma$), thus implying the presence of a degeneracy where {$0\leq r_1\leq 1$ and $0 \leq r_2 \leq 1$} may assume arbitrary values. This is indicated as a the vertical dashed line in Fig.~\ref{fig:Bifurcation_diagram_uncorrelated_model} (bottom).

Considering the numerical simulation results shown in Fig.~\ref{fig:RS_nocorr}, 
the Breathing \LLLL\ state exists inside a small region $1/2<p\lesssim 0.525=p_2$. This parameter region grows in size as $\gamma$ is increased, which is shown as the gray shaded region on the stability diagram in Fig.~\ref{fig:stability_diagram_uncorrelated_model}.

\subsubsection{Transverse stability of symmetric subspace \SS}\label{sec:Transverse_Stability_UncorrModel}
We so far only discussed stability on the symmetric (invariant) subset \SS\ with $z_1(t)=z_4(t)$ and $z_2(t)=z_3(t)$. It remains unclear whether or not the subset \SS\ is stable with respect to perturbations in directions transverse to itself, and in particular in the proximity of the \LLLL\ states. Unfortunately, deciding this question in general turns out to be cumbersome 
since the associated variational equations do not appear decouple in suitable directions.
However, numerical solutions of the governing equations~\eqref{eq:model} (see Figs.~\ref{fig:phi_uncorrelated} and Figs.~\ref{fig:simulation}) and the four complex Ott-Antonsen equations in $z_1,z_2,z_3,z_4$ (see \eqref{eq:zeq_uncorrelated_1}-\eqref{eq:zeq_uncorrelated_4} or Appendix~\ref{sec:appendix}) have confirmed stability for both Stable and Breathing \LLLL\ states in transverse direction of \SS, for all parameters we tested.

{For \INC\, the Ott-Antonsen equations~\eqref{eq:zeq_uncorrelated_1}-\eqref{eq:zeq_uncorrelated_4} yield four zero eigenvalues, and four eigenvalues that are either negative for $p<1/2$ and positive for $p>1/2$;  furthermore, for $p<1/2$, direct integration of Eqs.~\eqref{eq:zeq_uncorrelated_1}-\eqref{eq:zeq_uncorrelated_4} reveals a degeneracy with respect to \emph{random} initial conditions, as it is seen that $r_1,r_2,r_3,r_4$ converge to seemingly arbitrary values on $[0,1]$ as $t\rightarrow\infty $, rather than just 0, while $\phi_1(t)-\phi_4(t)\rightarrow 0$ and  $\phi_2(t)-\phi_3(t)\rightarrow 0$ as $t\rightarrow\infty $. }

\subsubsection{Stability diagram} 
\begin{figure}[htp!]
    \centering
    \includegraphics[width=.85\columnwidth]{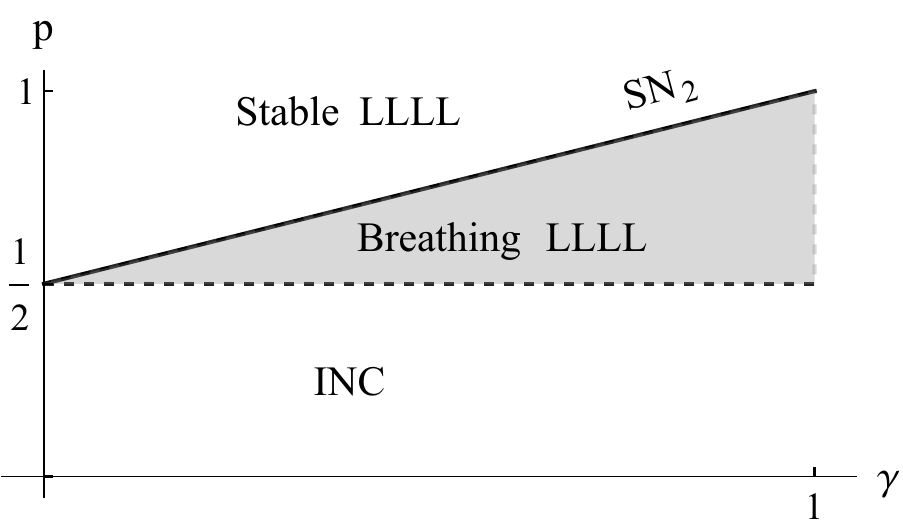}
    \caption{\label{fig:stability_diagram_uncorrelated_model}Stability diagram for the uncorrelated model in the continuum limit, $N\rightarrow\infty$. 
    }
\end{figure}

The preceding stability analysis for \INC\ and \LLLL\ states is summarized in the stability diagram shown in Fig.~\ref{fig:stability_diagram_uncorrelated_model}.
The dotted line delineates the stability boundary where \INC\ and Breathing \LLLL\ swap stability, see also Fig.~\ref{fig:Bifurcation_diagram_uncorrelated_model}. 
\subsubsection{Global order parameters}
On the symmetric subspace \SS, the global order parameters simplify to
\begin{align}
    R&=\sqrt{ p^2 r_1^2+2 p q r_1 r_2 \cos { \delta }+q^2 r_2^2}\\
    S&= \sqrt{(p-q)^2 \left(p^2 r_1^2+2 p q r_1 r_2 \cos { \delta }+q^2 r_2^2\right)}\
\end{align}
which for \INC\ ($0<p<1/2$) become $R=S=0$; and for the stable \LLLL\ state with $p\in[p_2,1]$ they become and $R=\sqrt{p^2+2pq\cos{\delta^*}+q^2}$ and $S=\sqrt{2p(p^2+2pq\cos{\delta^*}+q^2)}$, where $\delta^*=\arcsin{\gamma/(2p-1)}$. The breathing \LLLL\ state is bounded with $R_\text{min}:=2p-1\leq R\leq 1=:R_\text{max}$ and $(2p-1)^2<S<(2p-1)$ since then $|\cos{\delta(t)}|\leq 1$. 
This aligns with the phase diagram provided in Fig.~\ref{fig:RS_nocorr}, with the exception of two minor differences: 
(i) numerical simulations in the Stable \INC\ regime show that $R$ stays close to $R\approx 0$. Possible explanations for this behavior are manifold: finite size effects, critical slowing down near the bifurcation point $p_c=1/2$, and/or the aforementioned degeneracy of the \INC\ state; 
(ii) results in the Breathing \LLLL\ regime show values at the end of the simulation within the ranges specified above. 

{It is possible to determine an explicit expression for $S=S(p)$ in the stable \LLLL\ regime by deriving a self-consistency equation in the weighted order parameter 
~\eqref{eq:W_continuous} for the coherent (phase-locked) state based on Kuramoto's classical argument~\cite{Kuramoto1984,Strogatz2000}, see App.~\ref{sec:AppendixC}:
\begin{align}
    S&=\sqrt{8p^4-16p^3+14p^2+2\sqrt{A}-6p+1},
    \label{eq:selfcons_S_sol}
\end{align}
where $ A:= p^2(2p^2-3p+1)^2 (-\gamma^2+4p^2-4p+1) $. This result is numerically confirmed using numerical simulations, as shown in Fig.~\ref{fig:RS_nocorr}.
}

\section{Discussion}

\paragraph{Summary.}
We have studied the collective dynamics in a network of coupled {phase oscillators with disorder} in natural frequencies and coupling strengths, which were correlated or uncorrelated. Specifically, we have assumed that the coupling strength and the natural frequency of each oscillator may assume only one of two  values (positive or negative), amounting to a \emph{``Two-Frequency-Two-Coupling model''}. The character and stability of the nontrivial dynamic states in the models with correlated/uncorrelated disorder depend on the interplay of the disorder asymmetry parameter, $p$, and the frequency spacing, $\gamma$.
To explore how the different types of disorder influence the emergent phase coherence in the system, we performed numerical simulations revealing several nontrivial dynamic states. For the model with correlated disorder, oscillators split into two subpopulations where either one or both  subpopulations are perfectly phase-locked, amounting to Lock-Drift (\LD) or Lock-Lock (\LL) states, respectively. Both states maintain a constant phase difference, the size of which is controlled by the disorder asymmetry $p$.  \LD\ is stable for $p<p_c$ and swaps stability with \LL\ when $p>p_c$. This observation can be rationalized by observing that a majority of oscillators experience attractive ($\xi=1$) rather than repulsive coupling strength ($\xi=-1$) for large $p$. Equilibria for the global order parameters, i.e., the weighted $W$ and the unweighted $Z$, depend on $p$ in a nontrivial way (see Fig.~\ref{fig:R_S}). Furthermore, both states can be characterized by a traveling wave motion, corresponding to a non-zero mean-field frequency $ \Omega \neq 0$. At first sight,  the \LD\ state may resemble a chimera state, which also is characterized by one locked and one drifting subpopulation; however, the \LD\ state  is distinct since its symmetric counterpart, the \DL\ state, is unstable, thus reflecting that the asymmetry inherent to the system itself rather than the system dynamics gives rise to asymmetric states.
For the model with uncorrelated disorder, numerical simulations  indicated that oscillators split into four subpopulations all of which are phase-locked (Lock-Lock-Lock-Lock / \LLLL\ state); however, the two subpopulations with identical natural frequencies and opposing coupling strengths form pairs. These pairs are either frequency-locked  with constant phase difference (Stable \LLLL), or a drifting phase difference (Breathing \LLLL), see Fig.~\ref{fig:simulation}. For uncorrelated disorder, we also observed a state of Incoherence (\INC) where both global order parameters stay close to $R=S=0$. 

Next, we carried out a detailed bifurcation analysis for the local order parameters $z_\sigma(t)$ describing the collective dynamics and the synchronization level in subpopulations formed by oscillators with identical attributes (natural frequency / coupling strength). While uncorrelated disorder allows for the formation of $M=4$ subpopulations ($\Subpop_1,\Subpop_2,\Subpop_3,\Subpop_4$), correlated disorder naturally implies the presence of only $M=2$ subpopulations, $\Subpop_1$ and $\Subpop_2$, one with attractive coupling and negative frequency, the other with repulsive coupling and positive frequency.
The order parameters $z_\sigma $ satisfy Eqs.~\eqref{eq:z_dynamics} which can be derived using the Ott-Antonsen method~\cite{Ott2008,Bick2020} or the Watanabe-Strogatz method valid in the limit of $N\rightarrow\infty$ oscillators with uniformly distributed constants of motion. 

For correlated disorder, the \LD\ ($p<p_c(\gamma)$) and the \LL\ ($p>p_c(\gamma)$) states swap stability in a transcritical bifurcation at the critical value $p=p_c(\gamma) $ which we determined analytically (see Eq.~\eqref{eq:pcrit_gammacrit}). We also found analytical expressions of the unweighted and weighted order parameter $R=|Z|$ and $S=|W|$ at equilibrium, as well as for the non-zero mean-field frequency, $\Omega\neq 0$, thus prompting a traveling wave motion (Fig.~\ref{fig:R_S}). 

For uncorrelated disorder, the dynamics of locked states are confined to the (invariant) symmetric subspace \SS\, implying that subpopulations $\Subpop_1$ and $\Subpop_4$, and $\Subpop_2$ and $\Subpop_3$ are mutually phase-locked ($z_1(t)=z_4(t)$ and $z_2(t)=z_3(t)$). While a proof for stability transverse to the symmetric subspace \SS\ remained elusive, both numerical simulations of \eqref{eq:model} and direct numerical integration of \eqref{eq:zeq_uncorrelated_1}-\eqref{eq:zeq_uncorrelated_4} confirmed that \SS\ is attractive for two types of locked states.
The Stable \LLLL\ appears for $p>p_2$ and loses stability in a saddle-node bifurcation at $p=p_2$ on the invariant synchronized subspace defined by $r_1=r_2=1$; this gives rise to the Breathing \LLLL\ state which is stable for $1/2<p<p_2$. The Breathing \LLLL\ state is characterized by a drifting phase-relationship between subpopulations $\Subpop_1$ and $\Subpop_2$, i.e., their phase difference $\delta(t)$ increases monotonically and results in a periodic motion in $R(t)$ and $S(t)$.  
The Breathing \LLLL\ state is remarkable in the sense that there is no  external periodic driving acting on the system; i.e., the periodic synchronization emerges ``spontaneously'' when the coupling strengths and the natural frequencies are uncorrelated. Unlike for the correlated model, we did not find signs of traveling wave behavior with non-zero mean-field frequency $\Omega$; however, for the Breathing \LLLL\ state, two subpopulation pairs $(\Subpop_1,\Subpop_4)$ and $(\Subpop_2,\Subpop_3)$ drift apart while their average frequency stays close to 0 --- which we may refer to as a {\it ``Standing Wave''}, alike states observed for oscillator populations with bimodal frequency distributions~\cite{Crawford1994,Martens2009}. Both \LLLL\ states swap stability with the \INC\ state at $p=1/2$.

The Incoherent (\INC) state is always unstable in the correlated model, in contrast to the uncorrelated model, where \INC\ is neutrally stable for $p<1/2$ {on the symmetric subspace \SS}: Eqs.~\eqref{eq:z_dynamics} exhibit for the \INC\ state four negative and four zero eigenvalues for the \INC\ state. Numerical integration of Eqs.~\eqref{eq:z_dynamics} reveals degeneracy in the magnitude of local order parameters, i.e., $r_\sigma(t)$ may attain arbitrary stationary values between 0 and 1, which do not match Incoherence ($R=S=0$), while $|\phi_1-\phi_4|\rightarrow 0$ and $|\phi_2-\phi_3|\rightarrow 0$ as $t\rightarrow \infty$. {Therefore, we also expect that numerical simulations of Eqs.~\eqref{eq:model} may reveal states for $p<1/2 $ that have nonvanishing order parameters.} {However, introducing distributed frequencies of width $\Delta$ around each mode ($\omega=-q\gamma,p\gamma$) results in additional terms of the form $-\Delta z_\sigma $ in \eqref{eq:z_dynamics} for $\sigma=1,2,3,4$. This removes the degeneracy and renders \INC\ into a (stable) hyperbolic equilibrium, and we can say that \INC\ is a robust state. Establishing a complete bifurcation diagram for distributed frequencies is beyond the scope of this study and remains subject for future research.}
 
\paragraph{Relationship with other studies.}
The present study is closely related with previous work by Hong {\it et al.}~\cite{Hong2016} where oscillators' natural frequencies were drawn from a  distribution with finite nonzero variance (contrasting the zero-width distribution considered here), in order to explore the effects of symmetrically and asymmetrically correlated disorder.
It was found that asymmetrically correlated disorder induces traveling wave motion, characterized by non-zero mean-field frequency $\Omega\neq 0$; here, we found that correlated disorder still induces traveling waves when natural frequencies are bimodally distributed with zero variance, whereas uncorrelated disorder does not promote traveling waves. Thus, together with the simplifications implied by the present model, we may conclude that the traveling wave motion results from heterogeneity in terms of asymmetry in natural frequencies and coupling strengths, rather than it is a consequence of distributions with nonzero variance.
 
The correlated model also relates to several other studies addressing the collective dynamics of two interacting populations, either characterized by non-uniform interactions, see 
Abrams {\it et al.}~\cite{Abrams2008} and Martens {\it et al.}~\cite{Martens2016,Deschle2019,Bick2018};
the dynamics of populations with bimodal frequency distributions~\cite{Martens2009,Pazo2009,Crawford1994,Pietras2018};
or the dynamics of two population models combining both properties, see {\it Montbrío et al.}~\cite{Montbrio2004}, Laing~\cite{Laing2009} and Pietras~\cite{Pietras2016}.  
Most of these studies assume positive coupling strengths (exceptions include variants of the Kuramoto-Sakaguchi model with two populations~\cite{Martens2016} where heterogeneous phase-lags may result in negative coupling strength), whereas the correlated model has $W=p z_1 - qz_2$. One may interpret the prefactors $p$ and $-q$ in one of two ways: 
(i) in the generic way as the correlated model was posed, namely, natural frequencies are bimodally distributed with asymmetric peaks, populated by a fraction of oscillators $p$ and $q$ obeying attractive and repulsive coupled, respectively;
(ii) in the sense of (asymmetric) coupling strengths, i.e., when writing Eqs.\eqref{eq:z_dynamics} in matrix-vector notation with  the (vector) mean-field $W_c = \bigl( \begin{smallmatrix}p & -q\\ p & -q\end{smallmatrix}\bigr)
\cdot
\bigl( \begin{smallmatrix}z_1 \\ z_2\end{smallmatrix}\bigr)
$ promotes attractive (or excitatory) coupling with strength $p$ among oscillators within the first population and with the adjacent second population; and repulsive (or inhibitory) coupling $q$ among oscillators within the second population and the adjacent first population. 
The mean-field  for the uncorrelated model may also be interpreted in terms of coupling strengths in a similar fashion. Rewriting the mean-field in matrix-vector notation, we have 
$W_u = (2p-1)\bigl( \begin{smallmatrix}p & q\\ p & q\end{smallmatrix}\bigr)
\cdot
\bigl( \begin{smallmatrix}z_1 \\ z_2\end{smallmatrix}\bigr)$. Comparing Comparing $W_u$ with $W_c$ makes the different characters of the uncorrelated and the correlated model especially evident, as well as it elucidates why $p>\half$ or $p<\half$ results in predominantly attractive or repulsive coupling, promoting or hindering synchrony, respectively.

The models considered by Maistrenko {\it et al.}~\cite{Maistrenko2014} and Teichmann and Rosenblum~\cite{Teichmann2019} coincides with our model Eqs.~\eqref{eq:model}, but there are important differences. Our study concerns the effects of correlated/uncorrelated disorder on the long-term collective behavior and on their phase transitions towards synchrony for the thermodynamic limit ($N\rightarrow \infty$);
these authors studied solitary states in finite oscillator systems, where a single oscillator 'escapes' from the synchronized frequency cluster as repulsive interactions increase, however they disappear in the thermodynamic limit $N\rightarrow \infty$. 
Both models~\cite{Maistrenko2014,Teichmann2019} are restricted to subpopulations with equal size, $N_1=N_2=N/2$, corresponding to $p=1/2$ in our model for the thermodynamic limit; here, we studied the general case with $0\leq p \leq 1$.
Maistrenko {\it et al.} considered identical natural frequencies ($\gamma=0$), while 
Teichmann and Rosenblum~\cite{Teichmann2019}, considered the case with different natural frequencies in the subpopulations with attractive and repulsive (self-)interaction and found that the transition from a two-cluster synchrony to partial synchrony occurs via the formation of a solitary state for small frequency mismatch.

\paragraph{Outlook.}

Our analytical results are constrained to the dynamics on the Poisson manifold discovered by Ott and Antonsen~\cite{Ott2008,Marvel2009}; it would be interesting to investigate the dynamics off this manifold, too. Furthermore, it would be desirable to better understand how robust the Incoherent state in the correlated model is with regard to perturbations of the system.
The simplicity of the model suggests that real-world systems can be found that display the dynamic states  induced by correlated/uncorrelated disorder that we reported here. In this context it could be fruitful to identify intuitive mechanisms for, e.g., the breathing of the \LLLL\ state, generating periodic behavior of $R(t)$. Candidates for experimental systems might for instance be found in Josephson junction arrays~\cite{Wiesenfeld1998}, coupled Belousov-Zhabotinsky oscillators~\cite{Taylor2009,Tinsley2012,Totz2017,Calugaru2020}, and electro-chemical oscillators~\cite{Kiss2002,Wickramasinghe2013}.

\section{Acknowledgments}
This research was supported by the NRF Grant No.2021R1A2B5B01001951 (H.H). The authors thank C.~Bick for helpful conversations.

\section*{Data availability}
Data sharing is not applicable to this article as no new data were
created or analyzed in this study.

\appendix
\section{Full Ott-Antonsen equations for uncorrelated model\label{sec:appendix}}
For completeness, we list the Ott-Antonsen equations for the uncorrelated model in polar coordinates, describing the complete dynamics on the Poisson manifold (i.e., on and off the symmetric subspace \SS).
Rather than performing a bifurcation analysis for this system, we numerically solved the four ordinary differential equations above for the fixed point conditions $(\dot{r_1}=\dot{r_2}=\dot{r_3}=\dot{r_4}=0)$ for $p>p_c(=1/2)$. 
Introducing $z_\sigma \equiv r_\sigma e^{-i\theta_\sigma}, \sigma =1,2,3,4$,
we have
\begin{align}\label{eq:corr_r1} 
\begin{split}
    \dot{r}_1 &= \frac{1}{2} \Bigg[p^2 r_1 - p q r_4\cos\delta_{41} + p q r_3 \cos\delta_{31}\\
    &  - q^2 r_2 \cos\delta_{21}  -p^2 r_1^3 +pq r_4 r_1^2 \cos\delta_{41}\\
    &- p q r_3 r_1^2 \cos\delta_{31} + q^2 r_2 r_1^2 \cos\delta_{21}\Bigg], \
\end{split}    
\end{align}

\begin{align}\label{eq:corr_theta1}
\begin{split}
    \dot{\theta}_1 &=
    -q\gamma - \frac{1}{2r_1} 
    \Bigg[pqr_4 \sin\delta_{41} -pq r_3\sin\delta_{31} \\
    &+q^2 r_2 \sin\delta_{21} + pqr_4 r_1^2 \sin\delta_{41}\\
    &-pq r_3 r_1^2 \sin\delta_{31}+q^2 {r_2}r_1^2 \sin\delta_{21}\Bigg],\
\end{split}
\end{align}
where we defined the phase difference $\delta_{kl}(t):=\theta_k(t) - \theta_l(t)$ with $k$ and $l=1,2,3,4$.
Similarly, we find 
\begin{align}\label{eq:corr_r2}
    \begin{split}
    \dot{r}_2 &= 
    \frac{1}{2} \Bigg[p^2 r_1 \cos\delta_{21} -pq r_4\cos\delta_{42}
    +pq r_3 \cos\delta_{32} \\
    &-q^2 r_2-p^2 r_1 r_2^2 \cos\delta_{21}+pq r_4 r_2^2 \cos\delta_{42} \\
    &-pq r_3 r_2^2 \cos\delta_{32}+q^2 r_2^3\Bigg], 
    \end{split}
\end{align}

\begin{align} \label{eq:corr_theta2}
\begin{split}
    \dot{\theta}_2 &= 
    p\gamma - \frac{1}{2r_2} \Bigg[p^2r_1 \sin\delta_{21} + pq r_4\sin\delta_{42} \\
    &- pq r_3 \sin\delta_{32} + p^2r_1 r_2^2 \sin\delta_{21} \\
    &+ pq r_4 r_2^2 \sin\delta_{42}-pq {r_3}r_2^2 \sin\delta_{32}\Bigg],\
     \end{split}
\end{align}

\begin{align}\label{eq:corr_r3}
\begin{split}
    \dot{r}_3 &= 
    \frac{1}{2} \Bigg[p^2 r_1 \cos\delta_{31} -pq r_4\cos\delta_{43}
    + pq r_3 \\
    & - q^2 r_2 \cos\delta_{32} - p^2 r_1 r_3^2 \cos\delta_{31}+pq r_4 r_3^2 \cos\delta_{43} \\
    &-pq r_3^3+q^2 r_2 r_3^2 \cos\delta_{32}\Bigg], 
     \
\end{split}
\end{align}

\begin{align}\label{eq:corr_theta3} 
    \begin{split}
    \dot{\theta}_3 
    &= p\gamma - \frac{1}{2r_3} 
    \Bigg[p^2r_1 \sin\delta_{31} + pq r_4\sin\delta_{43} \\
    &-q^2 r_2 \sin\delta_{32} + p^2r_1 r_3^2 \sin\delta_{31}\\
    & +pq r_4 r_3^2 \sin\delta_{43}-q^2 {r_2}r_3^2 \sin\delta_{32}\Bigg], \
    \end{split}
\end{align}

\begin{align}\label{eq:corr_r4} 
\begin{split}
    \dot{r}_4 &= 
    \frac{1}{2} \Bigg[p^2 r_1 \cos\delta_{41} -pq r_4 +pq r_3 \cos\delta_{43} \\
    &-q^2 r_2 \cos\delta_{42} - p^2 r_1 r_4^2 \cos\delta_{41}+pq r_4^3 \\
    &-pq r_3 r_4^2 \cos\delta_{43} + q^2 r_2 r_4^2 \cos\delta_{42}\Bigg], \
\end{split}
\end{align}

\begin{align}
\label{eq:corr_theta4} 
\begin{split}    
    \dot{\theta}_4 &=
    -q\gamma - \frac{1}{2r_4}
    \Bigg[p^2r_1 \sin\delta_{41} +pq r_3\sin\delta_{43}\\
    &-q^2 r_2 \sin\delta_{42}  + p^2r_1 r_4^2 \sin\delta_{41}\\
    &+pq r_3 r_4^2 \sin\delta_{43}-q^2 {r_2}r_4^2 \sin\delta_{42}\Bigg]. \
\end{split}
\end{align}

With Eq.~(\ref{eq:corr_r1})-(\ref{eq:corr_theta4}), the order parameters $R = |Z|$ and $S = |W|$ are then given by
\begin{align}  \label{eq:R_uncorrelated_4D_OA}
\begin{split}
    R &=
    \Big[p^4 r_1^2 + p^2 q^2 r_4^2 +p^2 q^2 r_3^2 +q^4 r_2^2 \\
    &+2p^3 q r_1 r_4 \cos\delta_{41}+2 p q^3 r_3 r_2 \cos\delta_{32} \\
    &+2p^3 q r_1 r_3 \cos\delta_{31} +2p^2 q^2 r_1 r_2 \cos\delta_{21} \\
    &+2p^2 q^2 r_4 r_3 \cos\delta_{43}+2p q^3 r_4 r_2 \cos\delta_{42} \Big]^{1/2},~
\end{split}    
\end{align}  
and
\begin{align}   \label{eq:S_uncorrelated_4D_OA}  
\begin{split}
    S &=
    \Big[p^4 r_1^2 + p^2 q^2 r_4^2 +p^2 q^2 r_3^2 +q^4 r_2^2 \\
    &-2p^3 q r_1 r_4 \cos\delta_{41}-2 p q^3 r_3 r_2 \cos\delta_{32} \\
    &+2p^3 q r_1 r_3 \cos\delta_{31} -2p^2 q^2 r_1 r_2 \cos\delta_{21} \\
    &-2p^2 q^2 r_4 r_3 \cos\delta_{43}+2p q^3 r_4 r_2 \cos\delta_{42} \Big]^{1/2}.~
\end{split}
\end{align}  
Note that the $R$ and $S$ in Eq.~\eqref{eq:R_uncorrelated_4D_OA} and \eqref{eq:S_uncorrelated_4D_OA}
are valid only for $p>p_c$ since the fixed points solutions are available only for $p>p_c$.  Eqs.~\eqref{eq:R_uncorrelated_4D_OA} and \eqref{eq:S_uncorrelated_4D_OA} show a good agreement with the numerical simulation data, as seen in Fig.~\ref{fig:RS_nocorr}.

\section{Wave Speed}
\label{sec:AppendixB}
We derive the wave speed stated in Eq.~\eqref{eq:Omega}.
We have $W(t)=S(t)e^{i\Delta(t)}$ where $W, S ,\Delta \in \mathbb{R}$, and 
\begin{align}\label{eq:Wderivative}
 \dot{W} &= (\dot{S} +i S \dot{\Delta}) e^{i\Delta}, \
\end{align}
which implies $\dot{\Delta}=-\frac{i}{S}(\dot{W}e^{-i\Delta}-\dot{S})$.
However, since $\Delta \in \mathbb{R}$ we require $\dot{\Delta}\in\mathbb{R}$, and we therefore define
\begin{align}
 \Omega&:=|\dot{\Delta}|=\frac{1}{S}|\dot{W}e^{-i\Delta}-\dot{S}|.\
\end{align}
Observing the identity,
\begin{align}
 \dot{\bar{W}}e^{i\Delta}+\dot{W}e^{-i\Delta} &= 2\dot{S},\
\end{align}
obtained by substituting~\eqref{eq:Wderivative}, and $|z|^2 = z \bar{z}$, $z\in\mathbb{C}$, we obtain:
\begin{align}
 |\dot{W}e^{-i\Delta}-\dot{S}|^2 &= 
 |\dot{W}|^2 -\dot{S} (\dot{\bar{W}}e^{i\Delta}+\dot{W}e^{-i\Delta})+\dot{S}^2\\
 &= \sqrt{|\dot{W}|^2 - \dot{S}^2}. \
\end{align}
The wave speed is therefore
\begin{align}
 |\dot{\Delta}|&=\frac{1}{S}\sqrt{|\dot{W}|^2-\dot{S}^2}.
\end{align}

\section{Self-consistency argument for weighted order parameter}\label{sec:AppendixC}

The weighted order parameter in~\eqref{eq:W_continuous} allows us to re-cast the continuous version of the governing equations~\eqref{eq:goveq_continuous} into the following form,
\begin{equation}
    \dot\phi=\omega-S\sin(\phi-\Delta).
\end{equation}
We expect that a phase-locked solution with $\dot\phi=0$ with constant order parameter $W$ may exist when the effective coupling $S$ is sufficiently large to overcome the spread of the natural frequencies (i.e., when $S>|\omega|$). Accordingly, the locked phases are given by 
\begin{equation}
    \phi=\Delta+\sin^{-1}(\omega/S).
\end{equation}

In the continuum limit of $ N \rightarrow \infty $, the weighted order parameter \eqref{eq:W_continuous2} is expressed as follows:
\begin{align}\label{eq:selfconsistency}
    W=S e^{i\Delta} 
    &=\langle \xi \rangle \int e^{i(\Delta+\sin^{-1}(\omega/S))} g(\omega) \,\d\omega, \nonumber\\
    &= e^{i\Delta} \langle \xi \rangle \left[ \int_{-S}^{S} \sqrt{1-(\omega/S)^2} g(\omega)\, \d\omega \right],
\end{align}
where the stationary probability distribution function given by 
$\rho_{s}(\phi;\omega, \xi)=\delta(\phi-(\Delta+\sin^{-1}(\omega/S)))$, and $\langle \xi \rangle = \int \xi \,\Gamma(\xi)\,\d\xi=2p-1$ 
is the mean value of the distribution in $\xi$. 
Since oscillators assume either of two values for both coupling strengths and frequencies, we assume there is no contribution to the integral from drifting oscillators.
Carrying out the integral, we find that $S$ is implicitly given by the self-consistency equation
\begin{equation}
    S=(2p-1)[p\sqrt{1-(q\gamma/S)^2}+q\sqrt{1-(p\gamma/S)^2}].
    \label{eq:sc_S}
\end{equation}
We note that $S$ in Eq.~(\ref{eq:sc_S}) must satisfy the conditions
$S\geq q\gamma$ and $S \geq p\gamma$ in order to be real-valued; in other words, the interval of $p$ is restricted for which the self-consistency equation produces real values for $S$.

Substituting $q=1-p$, Eq.~(\ref{eq:sc_S}) we can find an exact solution for $S$ using algebraic manipulation software
given by
\begin{align}
    S&=\sqrt{8p^4-16p^3+14p^2+2\sqrt{A}-6p+1},
\end{align}
where $ A:= p^2(2p^2-3p+1)^2 (-\gamma^2+4p^2-4p+1) $.
This result is numerically confirmed using numerical simulations (Fig.~\ref{fig:RS_nocorr}).

\bibliographystyle{unsrt}

\end{document}